\documentclass[11pt]{article}

\usepackage[margin=1in]{geometry}
\usepackage[T1]{fontenc}
\usepackage[utf8]{inputenc}
\usepackage{lmodern}
\usepackage{microtype}

\usepackage{amsmath,amssymb,amsthm,mathtools}

\usepackage{graphicx}
\usepackage{booktabs,longtable,array,multirow}
\usepackage{float}
\usepackage{caption}
\usepackage{subcaption}

\usepackage{enumitem}

\usepackage[dvipsnames]{xcolor}
\usepackage{xurl}
\usepackage[colorlinks=true, linkcolor=blue, citecolor=blue, urlcolor=blue]{hyperref}
\usepackage[authoryear,round]{natbib}

\newenvironment{BlueText}{}{}
\newtheorem{theorem}{Theorem}

\def\spacingset#1{\renewcommand{\baselinestretch}{#1}\small\normalsize}

\def\B0{\mbox{\boldmath $0$}}
\def\Ba{\mbox{\boldmath $A$}}

\def\btheta{\mbox{\boldmath $\theta$}}
\def\htheta{\mbox{$\hat{\theta}$}}

\hypersetup{
  pdftitle={ETZ: A Modeling Principle for Confirmability of Drug-Development Studies},
  pdfauthor={Yujia Sun; Yang Han; Xingya Wang; Haiyan Xu; Szu-Yu Tang; Yushi Liu; Jason C. Hsu},
  pdfkeywords={Variability Decomposition, Randomized Controlled Trials, Decision-making Process, Multiple Comparison},
  colorlinks=true,
  linkcolor=blue,
  citecolor=blue,
  urlcolor=blue
}

\begin{document}

\title{\bf ETZ: A Modeling Principle for Confirmability of Drug-Development Studies}
\author{Yujia Sun\\
Department of Mathematics, The University of Manchester;\\
Dept. Data Science \& Big Data Technology, Zhejiang A\&F University\\
Yang Han*\\
Department of Mathematics, The University of Manchester\\
Xingya Wang\\
Department of Mathematics, The University of Manchester\\
Szu-Yu Tang\\
Pfizer Worldwide Research \& Development\\
Yushi Liu\\
Eli Lilly and Company\\
and\\
Jason C. Hsu*\\
Department of Statistics, The Ohio State University}
\date{}
\maketitle

\begin{abstract}
Transitioning from Phase 2 to Phase 3 in drug development, at a rate of $\approx$40\%, is the most stringent among phase transitions (\cite{HayEtAl(2014)}). Yet, success rate at Phase 3 leading to approval is only $\approx$50\% (\cite{Arrowsmith(2011)PhaseIII}). To improve \emph{Confirmability}, we propose a methodological shift: replacing \textit{multiple hypothesis testing} with inference based on confidence sets, and substituting conventional \textit{power and sample size} calculations with a Confidently Bounded Quantile (CBQ) framework.

Our confidence set inferences to answer the questions of whether to transition to a Confirmatory study as well as what to designate as the endpoint in that study. Construction of our \textit{directed} confidence sets follows the Partitioning Principle, taking the best of each of Pivoting and Neyman Confidence Set Construction.

Rooted in Tukey's Confidently Bounded Allowance (CBA) (\cite{Tukey(1994)Chapter18}), our proposed CBQ makes the transitioning decision following the Correct and Useful Inference principle in \cite{Hsu(1996)}. CBQ removes from ``power'' the probability of \emph{rejecting for wrong reasons}, eliminating the need for informal \textit{discounting} in power calculation that has existed in the biopharmaceutical industry.

ETZ, the modeling principle proposed in \cite{wang2025counterfactual}, quantifies the impact of three variability components on \textit{confirmability}. In repeated-measures RCTs, it separates \textit{within-subject} and \textit{between-subject} variability, further dividing the latter into \textit{baseline} and \textit{trajectory} components. This enables informed investment decisions for the sponsors on targeting variability reduction to improve \textit{confirmability}. A Shiny-based \textit{Confirmability App} supports all computations.
\end{abstract}


\spacingset{1.8}

\begin{figure}[H]
\centering
\includegraphics[width=0.8\linewidth]{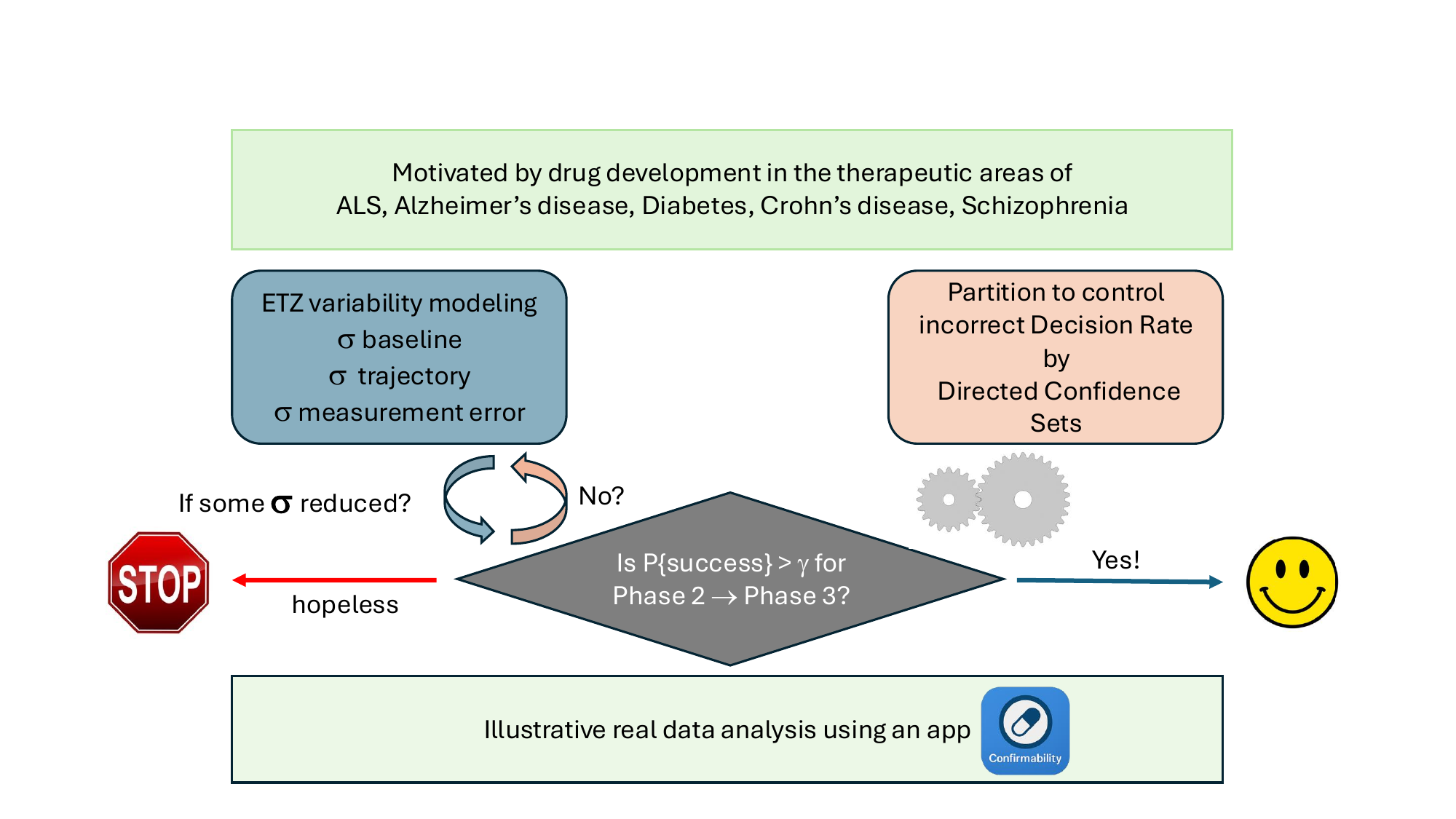}
\caption{The structure of the article.}
\label{fig:PaperStructure}
\end{figure}

\section{Lack of confirmability of drug development studies}
Drug development proceeds in stages, from Phase 1 through Phase 3 to New Drug Application (NDA) or New Biologics License Application (BLA). Among the Phase transitions, the Phase 2 to Phase 3 transition rate at 39\% as reported by \cite{HayEtAl(2014)} is the most stringent, with rates of 67\% for Phase 1 to 2, 68\% for Phase 3 to NDA/BLA, 86\% from NDA/BLA to approval. Yet, the success rate at Phase 3 leading to approval is only $\approx$50\% as reported by \cite{Arrowsmith(2011)PhaseIII}.

This fact seems puzzling since only Phase 2 studies with promising results graduate to Phase 3 studies, and Phase 3 studies have large sample sizes typically calculated to achieve ``power'' of 80\%, 90\%, or even higher. Within late Phase studies, we aim to increase the Phase transition success rate, so that a promising Phase 2 result will have a high probability of being confirmed in a Phase 3 study, for example.

For simplicity, we phrase the setting of this article as comparing a new treatment $ Rx $ with a control treatment $ C $ (a placebo or a standard-of-care). Our \textit{reduce-variability principle} for enhancing the rate of success transitioning from a \textit{feeder} study to a \textit{confirmatory} study has the following decision path:

\begin{description}[leftmargin=1.2em]
\item[Misleading feeder inference?] Learn, from the feeder study data itself, whether the promising result may be selective from highly variable potential results;
\begin{description}[leftmargin=1.2em]
\item[No] Proceed to a confirmatory study.
\item[Yes] Go to whether reducing variability is worthwhile checking below.
\end{description}
\item[Variability reducible?] Can ETZ variability components be sufficiently reduced to achieve high confidence in confirmability?
\begin{description}[leftmargin=1.2em]
\item[No] Terminate development.
\item[Yes] Allocate resources to reduce impactful ETZ variabilities, then proceed.
\end{description}
\end{description}

Drug development studies are typically randomized controlled trials (RCTs), so unmeasured confounders that might mislead observational studies are unlikely to be the causes of failure. For studies with repeated measures, we have a way of decomposing variability into three clinically meaningful components (abbreviated as $ E $, $ Traj $, and $ Z $), beyond just \textit{between-} and \textit{within-} patient variabilities. We also have a way of assessing separately to what extent each component can cause a failure to confirm, so that resources can be judiciously allocated to reduce the most impactful component. A key innovation of our research is to show ETZ variabilities can be estimated from three variances routinely reported in RCTs.

\paragraph*{Confirmability as distinct from reproducibility and replicability}
We follow the language in \cite{NationalAcademy(2019)} for \textit{reproducibility} and \textit{replicability}, adding the terminology of \textit{confirmability} for clarity.

\textit{Reproducibility} means that independent investigators can reproduce the reported figures, tables, point estimates, confidence intervals, etc., of a specific study from its original data, performing the same or equivalent analyses, using the same or equivalent computer codes. Reproducibility Research checks are exercised by many journals before publication for reproducibility purposes.

In the context of this article, we interpret \textit{replicability} as the \textit{result} of a study that can be essentially replicated by a similarly study (similar in design including sample sizes, comparing the same or similar class of compounds, toward the same or similar indications). Replicability would apply within each Phase of an RCT. \cite{Arrowsmith(2011)PhaseII} showed the overall success rate of Phase 2 studies is less than 20\%, and \cite{Arrowsmith(2011)PhaseIII} showed approximately 50\% of Phase 3 drug development trials fail, so replicability rate may differ by Phases of drug development.

\paragraph*{Dissimilarities between a feeder study and a confirmatory study}
It is an oversimplification to think of a Phase 3 confirmatory study as like its Phase 2 feeder with a larger sample size.

Either a Phase 2 study with promising result, or a Phase 3 study which fails its original objective but shows tantalizing possibilities for the compound, can be a feeder to a ``confirmatory'' Phase 3 study. Besides sample sizes, common examples of how feeder and confirmatory studies differ are as follows.

In transitioning from a promising dose-finding Phase 2 study to a Phase 3 confirmatory study, often only the most promising dose is selected. In transitioning from a Phase 3 study which fails to show efficacy in the primary endpoint but appears to have efficacy in a key secondary endpoint, the roles of primary and secondary endpoints may be reversed in a subsequent Phase 3 confirmatory study. In both cases, if a subgroup of patients appears to obtain higher efficacy than the overall population in the feeder study, it is common for the subsequent confirmatory study to primarily target that subgroup of patients.

A study fails if, at the end of it, no clinically meaningful separation between the effects of $ Rx $ and $ C $ is observed. \textit{High variability} in the observed endpoint separation of $ Rx $ and $ C $, is the root cause of the low rate of confirmability. Our ability to quantify how reducing each ETZ variability component can increase the chance of successful confirmation allows the drug developer to decide whether to invest in additional resources to reduce the variability of a particularly impactful component. Our research is toward increasing the \textit{confirmability} of study results.

\subsection{Therapeutic areas with Repeated Measures studies}
It takes at least two measurements to estimate a variability component, so traditional thinking is to separate between-treatment, between-patient, and within-patient variabilities. Ideally, a crossover study should be conducted, so each patient is given each treatment at least twice. However, in typical comparative efficacy clinical trials, each patient is given only one treatment.

The setting of this article is randomized controlled trials (RCTs) in which each patient is given \textit{either} $ Rx $ or $ C $ but not both, and treatment outcomes are measured \textit{repeatedly} at the \textit{same} time points on each subject, at Week 0, Week 4, $ \ldots $, Week 80 for example. Randomization and repeated measures allow us to separate the variability components, as we will show.

Therapeutic areas with randomized controlled trials that repeatedly measure patient outcome over time include diabetes such as type 2 diabetes mellitus (T2DM), neurodegenerative diseases such as Alzheimer's Disease (AD), psychiatric diseases such as schizophrenia, immunological diseases such as rheumatoid arthritis, ulcerative colitis, Crohn's disease, and psoriasis. Toward more confirmable drug development studies, Table \ref{table:NofPatients} gives an idea of the approximate number of patients that may benefit from insights given by our research.

\begin{table}[H]
\centering
\begin{tabular}{|c|c|c|}
\hline
Therapeutic Area & Number of Patients (in millions) & Region \\
\hline
Crohn’s disease and ulcerative colitis & 3 & U.S. \\
\hline
Rheumatoid arthritis & 54 & U.S. \\
\hline
Schizophrenia & 20 & worldwide \\
\hline
Alzheimer’s disease & 50 & worldwide \\
\hline
Type 2 diabetes & 400 & worldwide \\
\hline
\end{tabular}
\caption{Possible therapeutic areas influenced by our research.}
\label{table:NofPatients}
\end{table}

Typically in such studies, the measure of treatment effect is \textit{change from baseline}, the \textit{difference} in patient outcome between the milestone visit and the first visit. Milestone visit is the visit when primary efficacy analysis is taken, while the first visit (baseline) is the only visit when patients are not given any treatment. In most cases, the primary efficacy measure of $ Rx $ versus $ C $ is the expected \textit{difference} of the primary outcome measure (the difference of long run change-from-baseline averages).

At a high level, one can conceptualize that there is \emph{between} patient variability and a \textit{within} patient variability. There are actually two components of \emph{between} patient variability, as follows. At entry to a clinical trial, before any treatment is given, there is patient-to-patient variability (which we will later call \textit{intercept} variability). Once treatments are given, then within each treatment arm there is variability in how patients progress over time (which we will later call \textit{trajectory} variability). There is also \textit{within} patient variability, from random errors in measuring treatment outcome on each patient. We will refer this variability to \textit{measurement error} variability. Two examples illustrate how these variabilities can be big or small.

In an efficacy Type 2 diabetes mellitus (T2DM) study, the primary treatment outcome measure is glycated haemoglobin A1c (HbA1c). HbA1c measures the percent of red blood cells in the patient's blood with glucose attached to them, averaged over the last 3 months. HbA1c's change-from-baseline is a \emph{validated} surrogate endpoint for reduction of microvascular complications associated with diabetes mellitus. See \cite{BEST(2016)}. One reason for using HbA1c (as opposed to measuring microvascular complications as outcomes) is its relatively small measurement error, with a reported standard deviation (SD) of 0.17. A Trulicity$^{\textsuperscript{\textregistered}}$ study reported patients with entry HbA1c between 6.5 and 9.5 (that being the patient entry criterion) had a standard deviation (SD) of baseline HbA1c about 1.1. Even though this baseline variability includes both measurement error variability and intercept variability (which will be separated later in the article), nevertheless at this point of the article, one can sense for T2DM measurement error variability (with a SD of 0.17) is small compared to the between-patient intercept variability.

While diabetes has been treated for more than a century, developing \textit{disease-modifying} treatments for Alzheimer's disease (AD) remains challenging. In AD studies, patient outcomes are often their cognitive (Cog) abilities as measured on the Alzheimer’s Disease Assessment Scale-Cognitive (ADAS-Cog) scale, and abilities to perform instrumental activities of daily living (iADL) as measured on the Alzheimer Disease Cooperative Study-Instrumental Activities of Daily Living (ADCS-iADL) scale. ADAS-Cog11, for example, test 11 cognition items such as Word Recall, Naming Objects, etc. Treatment effects are commonly assessed as change-from-baseline of ADAS-Cog and ADCS-iADL. Scored by raters, a patient's ADAS-Cog measurement may be affected by whether they are having a good day or a bad day, and by rater variability.

For the EXPEDITION3 study (\cite{honig2018trial}) testing solanezumab for treating Alzheimer's disease, at the first visit, mean baseline ADCS-iADL was about 45 with a standard deviation (SD) of approximately 8. Later in this article we estimate SD of measurement error of ADCS-iADL to be about 3.3, an indication that for AD measurement error variability is not small compared to the between-patient intercept variability.

\subsubsection{Strategies for reducing variability}\label{sec:ReducingVar}
We describe four potential ways of reducing variability, with examples.

\begin{description}[leftmargin=1.2em]
\item[Narrow Entry] Narrow the patient entry criterion to reduce patient heterogeneity
\begin{itemize}
\item Type 2 diabetes: patients with HbA1c $ \in (7.0, 9.0) $ instead of HbA1c $ \in (6.5, 9.5) $
\end{itemize}
\item[Rater Training] Use more highly trained raters to reduce measurement error variability
\begin{itemize}
\item Schizophrenia: rate PANSS by raters at a centralized site
\end{itemize}
\item[Surrogate Endpoint] Use a surrogate endpoint less variable than clinical outcome
\begin{itemize}
\item Type 2 diabetes mellitus: measure HbA1c in the blood
\end{itemize}
\item[Subgroup Targeting] Enroll only patients with sufficient drug target expression
\begin{itemize}
\item Alzheimer's disease: patients with amyloid beta deposition (by PET imaging)
\end{itemize}
\end{description}

Imaging is costly, and so is having ratings done by highly trained raters at central locations (as opposed to by local clinicians or pathologists at trial centers). Surrogacy of biomarker-based endpoints also needs to be validated initially by large-scale studies that measure both the biomarker endpoint and clinical outcomes side-by-side. With finite resources to allocate, the questions are ``How much does each variability component contribute to a potential failure to confirm efficacy? Is it worth the expense of reducing it?'' The ETZ decomposition of variability components (to be developed in Section \ref{sec:EZdecomposition} of this article) provides a principled approach of assessing how much each component of variability contributes to a potential failure to confirm. It provides a rational way to design a confirmability study for success.

To concretely motivate our approach, we describe in the next section the Phase 3 EXPEDITION3 Alzheimer's disease study.

\subsubsection{EXPEDITION3 studying solanezumab for treating Alzheimer's disease}\label{sec:EXPEDITION3}
As reported in \cite{honig2018trial}, EXPEDITION3 studied solanezumab for treating patients with mild dementia due to Alzheimer’s disease. A double-blind, placebo-controlled, Phase 3 trial, EXPEDITION3 involved a total of 2129 patients randomized to receive either solanezumab at a dose of 400 mg or a placebo, every 4 weeks for 76 weeks (1057 patients were assigned to receive solanezumab and 1072 to receive placebo). Scores at the initial (Week 0) visit were considered \textit{baseline}, and thereafter treatment outcomes were measured at weeks 12, 28, 40, 52, 64, and 80.

\begin{figure}[H]
\centering
\includegraphics[height=0.65\textheight,keepaspectratio]{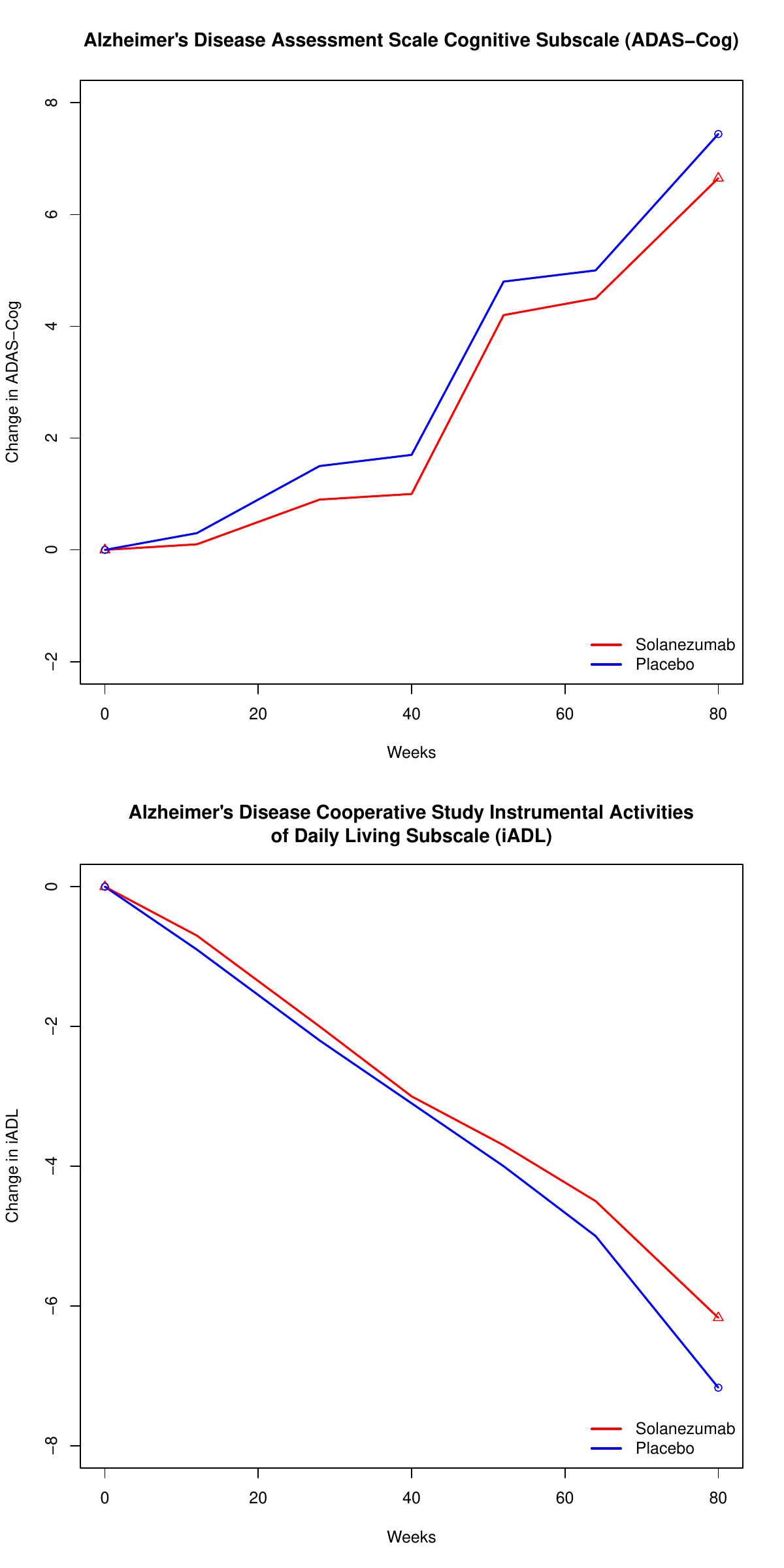}
\caption{Observed profiles of solanezumab vs. placebo for ADAS-Cog14 and ADCS-iADL}
\label{fig:SolanezumabProfiles}
\end{figure}

The primary outcome was the change (from baseline to Week 80) in the score on the 14-item cognitive subscale of the Alzheimer’s Disease Assessment Scale (ADAS-cog14), with higher scores indicating greater cognition loss. A key secondary endpoint was the change (from baseline to Week 80) of Alzheimer Disease Cooperative Study-Instrumental Activities of Daily Living (ADCS-iADL), which assesses complex activities such as using public transportation, managing finances, or shopping, with lower scores indicating greater functional loss. Figure \ref{fig:SolanezumabProfiles} shows observed profiles of ADAS-cog14 and ADCS-iADL reproduced from Figure 2 in \cite{honig2018trial}. EXPEDITION3 did not demonstrate efficacy in the chosen primary endpoint ADAS-Cog14, but did show promising results in the secondary endpoint ADCS-iADL. Planning of EXPEDITION3 was done according to standard power and sample size calculation techniques. What this article shows is that there are under-recognized factors that make it difficult to judge which promising results in earlier Phase or in the same Phase studies can be confirmed, and ETZ model in Section \ref{sec:EZdecomposition} will make under-recognized factors precise, so that they can potentially be mitigated.

For illustration purposes, since EXPEDITION3 indicated solanezumab may have potential in slowing the decline in ADCS-iADL (with a confidence interval for difference of change from baseline between solanezumab and the placebo being (0.17, 1.83)), we use the ADCS-iADL outcome in EXPEDITION3 to demonstrate the potential of our methodology to discern whether such observations can be confirmed in subsequent studies, or is an illusion due to variability, be it person-to-person variability or measurement error.

Analyses in EXPEDITION3 were by mixed effects repeated measures modeling, with visits considered a categorical variable. Our methodology can, surprisingly, extract sufficient information from such routine analyses, to discern how variable future studies will be, under a variety of variability conditions.

\subsection{Mixed Model Repeated Measures (MMRM)}
Index each of the $ m $ visits of a repeated measures study by $ v $, $ v = 1, \ldots, m$. Denote by $ Y_{i}^{Trt[v]} $ the outcome of a patient $ i $ given treatment $ Trt $ (either $ Rx \mbox{ or } C $) measured during visit $ v $.

Denote the \textit{time from randomization} of each subject as \textit{Time}, which is a common factor for all the subjects in an RCT. There are two ways of modeling the repeated measurements. The first, a random coefficient model, to be discussed in Section \ref{sec:RandomCoefficient}, considers Time as a continuous variable, with the response of each subject generated randomly from a distribution of response trajectories around parametric response profiles representing average effects of treatments $ Rx $ and $ C $ over Time.

The second model is called Mixed Model Repeated Measures (MMRM)
\begin{eqnarray}\label{model:MMRM}
Y_{i}^{Trt[v]} = \mu + \alpha^{[v]} + \beta^{Trt} + (\alpha\beta)^{Trt[v]} + b_{i}^{Trt} + w_{i}^{Trt[v]}, ~ v = 1, \ldots, m.
\end{eqnarray}
The fixed effect is represented by $ \mu + \alpha^{[v]} + \beta^{Trt} + (\alpha\beta)^{Trt[v]} $, the mean response under treatment $ Trt $ at time $ v $, where $ \alpha^{[v]} $ is the time effect, $ \beta^{Trt} $ is the treatment effect, and $ (\alpha\beta)^{Trt[v]} $ is the treatment$ \times $time interaction effect. Random between subject effect for the $ i^{th} $ patient assigned to treatment $ Trt$ is $ b_{i}^{Trt} $. Random within subject effect is represented by $ w_{i}^{Trt[v]} $.

In a repeated measures model, corresponding to the REPEATED statement in Proc Mixed of SAS, Time is not considered a continuous variable but rather just a label, a \texttt{CLASS} variable in SAS, or a \texttt{factor} in R. No profile of the response over time is modeled, but the \texttt{Type =} statement let the user specify a variety of structures of the $ m \times m $ variance-covariance matrix of measurements on the subjects across the $ m $ visits. This specification does not separate variability from the randomness of the individuals' response trajectories and the randomness of measurement errors.

MMRM has become popular, from recommendations such as
\begin{quote}
MMRM analysis appears to be a superior approach in controlling Type I error rates and minimizing biases, as compared to LOCF ANCOVA analysis (\cite{SiddiquiHung(2009)}).
\end{quote}

\section{The ETZ modeling principle}\label{sec:EZdecomposition}
\begin{BlueText}
Index each of the $ m $ visits of a repeated measures study by $ v $, $ v = 1, \ldots, m$. Denote by $ Y_{i}^{Trt[v]} $ the outcome of a patient given treatment $ Trt $ (either $ Rx \mbox{ or } C $) measured during visit $ v $.

The ETZ modeling principle thinks of each patient as having their own random \textit{intercept} before any treatment is given, with their random response \textit{trajectory} around a \textit{profile} of long-run average trajectories (be it linear, quadratic, Emax, what have you) once treatments are given. Values of profiles for $ Rx $ and $ C $ at times of visits are the fixed effects in the MMRM model (\ref{model:MMRM}). Similar to the MMRM model, ETZ does not assume a functional form for the profiles. Conceptually drawn as the curves in Figure \ref{fig:ETZ3RCT}, trajectories are not actually observed because measurements have errors. Observed, or potentially observed, are $ Y_{i}^{Trt[v]}, ~v = 1, \ldots, m $, represented as dots and squares in Figure \ref{fig:ETZ3RCT}.

\subsection{ETZ notations}
To be more specific, conceptualize $ Y^{Trt[v]} $ as having either two or three random components. At visit 1 ($ v = 1 $), before any treatment is given, $ Y^{Trt[1]} $ has two random components, a component $ Z $ which is that patient's intercept (true baseline measurement), and a measurement error component $ E $ which is the difference between the true and the observed baseline. Thereafter, $ Y^{Trt[v]} $ has three random components, a $ Z $ which is that patient's intercept (true baseline measurement), plus a random trajectory $ Traj $ component (around its mean profile $ \mu $) representing that patient's response to the treatment given up to the time of visit $ v $ ($ v > 1 $), and a measurement error component $ E $ which is the difference between the measured outcome and that patient's true trajectory at each visit $ v $. These components can be seen in Figure \ref{fig:ETZ3RCT}. Distinct from MMRM (\ref{model:MMRM}), the ETZ modeling principle clinically separates measurements $ Y^{Trt[1]} $ at visit 1 ($ v = 1 $) before any treatment is given from measurements $ Y^{Trt[v]} $ at visits $ v > 1 $ once treatments are given. See (\ref{eq:Ybefore}) and (\ref{eq:Yafter}) below. It is assumed that measurement error $ E $ is independent of $ Z $ and $ Traj $, and i.i.d. over the visits, across all the patients.

\begin{figure}[H]
\centering
\includegraphics[width=0.6\linewidth]{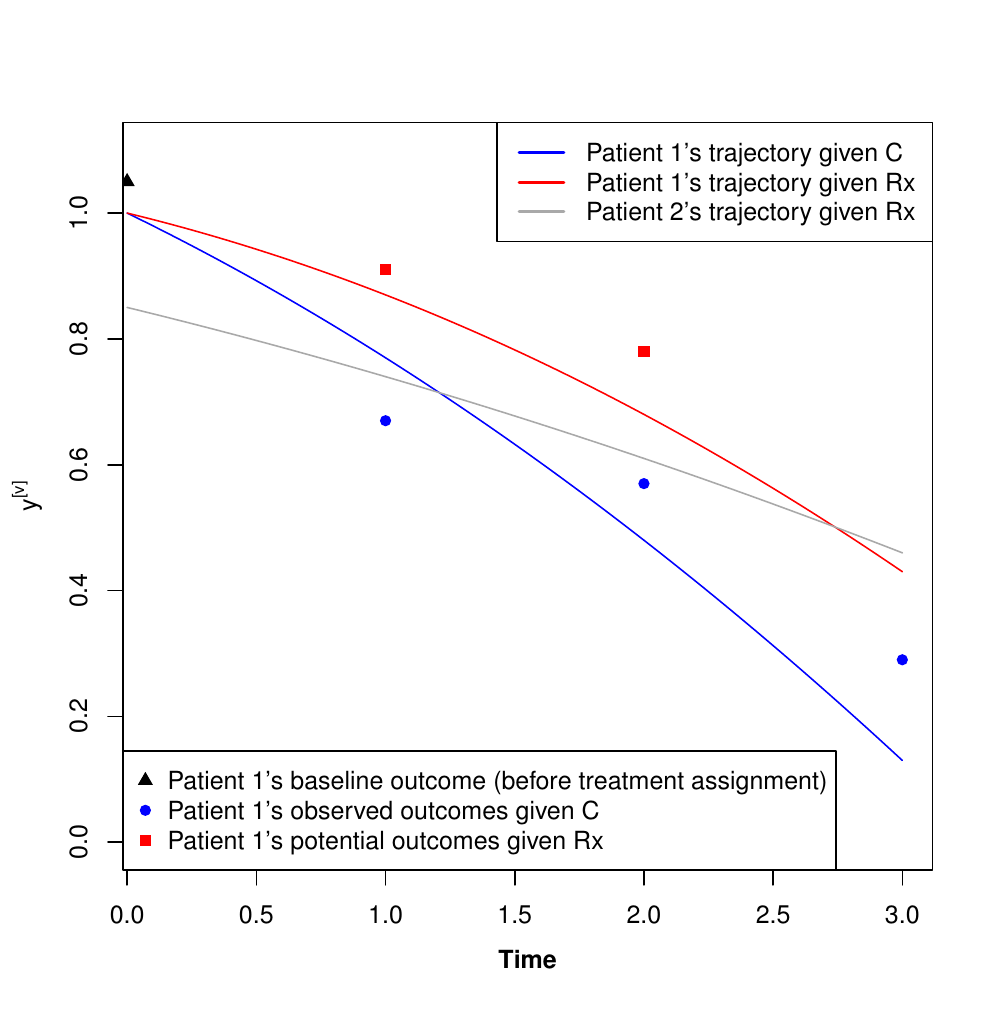}
\caption{Components of the ETZ modeling principle for an RCT with $ m = 4$ visits at Time = 0, 1, 2, and 3. The ETZ modeling principle only needs variability information for (1) baseline, (2) trajectory, and (3) milestone visits. No shape of the curve needs to be assumed, as Change-from-baseline sufficiently captures the trajectory information. The black triangle is the baseline of patient 1 before any treatment is taken. The blue curve is the trajectory of patient 1 who is assigned to $ C $, and the blue dots are the observed $ Y $. The red curve represents patient 1's potential trajectory had they been assigned to $ Rx $, and the red squares are the potential $ Y $ (with the last value missing). The grey curve is the trajectory of patient 2 who is assigned to $ Rx $. Measurement errors are differences between the outcomes and their trajectories.}
\label{fig:ETZ3RCT}
\end{figure}

\paragraph*{Variabilities reported in clinical trials}
With random allocation of patients to $ Rx $ and $ C $ in a randomized controlled trial (RCT), one can estimate unbiasedly $ Rx $ versus $ C $ effect by comparing observed patient outcomes under $ Rx $ and $ C $. Primary measure of treatment effect typically is \textit{change} from baseline, $Y^{Trt[\mathrm{change}]} =  Y^{Trt[m]} -Y^{Trt[1]}$, so reducing $ Var(Y^{Trt[\mathrm{change}]}) $ increases the chance of a successful confirmatory trial. A published result of an RCT typically includes an estimate of $ Var(Y^{Trt[1]}) $, and the standard error of the estimated mean of $ Y^{[\mathrm{change}]} $ from which one can infer an estimate of $ Var(Y^{Trt[\mathrm{change}]}) $. Often an estimate of $ Var(Y^{Trt[m]}) $ is given as well.

It is not obvious how these variabilities relate to each other. For example, would narrowing the patient entry criterion, which presumably reduces $ Var(Y^{Trt[1]}) $, have much of an impact on $ Var(Y^{Trt[\mathrm{change}]}) $? If it did, then narrowing patient entry criterion would increase the chance that a confirmatory study would be successful. It turns out that it actually does not, as we will show in Section \ref{sec:TrulicityEntry}.

\paragraph*{Idea}
The $ Traj $ (T) and $ Z $ parts of ETZ separate the between-subject effect $ b_{i}^{Trt} $ in the MMRM model (\ref{model:MMRM}) into two parts: a visit 1 pre-treatment part, and a treatment-dependent $ Traj $ part starting with visit 2. The measurement error $ E $ part of ETZ is different from the within-subject measurement error in the model (\ref{model:MMRM}). Each of $ E $, $ Traj $, and $ Z $ is medically interpretable, and their variabilities can be controlled to varying extents. For example, while intercept variability $ Var(Z) $ can be reduced by narrowing the patient entry criterion, trajectory variability $ Var(Traj) $ can be reduced by decreasing drug target heterogeneity among patients. Variability in measurement error $ Var(E) $ can be reduced by using more highly trained raters, for example.

High variability of change-of-baseline separation between $ Rx $ and $ C $ causes a low rate of successful confirmatory studies. How much of an impact on $ Var(Y^{Trt[\mathrm{change}]}) $ will reduce each of $ Var(Z) $ , $ Var(Traj) $, $ Var(E) $ have? That question will be answered in Section \ref{sec:ComponentImpacts}, after we show the three variances typically reported are sufficient to estimate variabilities of $ Z $, $ Traj $, and $ E $.

\paragraph*{Plausibility}
Intuitively, variability at the first and milestone visits and of Change may contain all the information we need to separate the $ E $, $ Traj $, and $ Z $ variability components. Baseline measurements at the beginning of the study (visit 1) have the random intercept $ Z $ and measurement error $ E $ components but no random trajectory component $ Traj $ since patients have not been treated yet. Change, or change-from-baseline, measurement at the milestone visit minus measurement at the first visit, has the random trajectory $ Traj $ and the measurement error $ E $ components (two of $ E $ in fact) but not the random intercept component since that has been subtracted. Measurements at the end of the study (visit $ m $) have all three random components. So it seems plausible that $ Var(Y^{Trt[1]}) $, $ Var(Y^{Trt[m]}) $, and $ Var(Y^{Trt[\mathrm{change}]}) $ are sufficient for us to recover the $ E $, $ Traj $, and $ Z $ variability components (assuming variability of measurement errors is constant across visits).

We assume the random intercept $ Z $ and the random trajectory $ Traj $ are \textit{independent} within each treatment arm, that is, a patient's true severity of illness before they receive either $ Rx $ or $ C $ does not inform on whether that patient's response trajectory is going to be above or below the profile $ \mu $ (long-run average of the trajectories). In addition, while trajectories for $ Rx $ and $ C $ are expected to have different profiles, we assume \textit{variabilities} of the trajectories are the same for $ Rx $ and $ C $.

\subsection{Components of the ETZ decomposition}
Assuming that the variability of measurement error at visit $ m $ is the same as the variability of measurement error at visit $ 1 $, our discussion starts from the setting in which random intercept $ Z $ is independent of the random trajectory $ Traj $.

With such a conceptualization, outcomes of patient $ r $ given $ Rx $ and patient $ s $ given $ C $ at visit 1 and visit $ m $ are represented as
\begin{eqnarray}
Y^{Rx[1]}_{r} & = & Z_{r} + E^{[1]}_{r} \label{eq:Ybefore}\\
Y^{Rx[m]}_{r} & = & Z_{r} + Traj^{Rx[m]}_{r} + E^{[m]}_{r} \label{eq:Yafter}\\
Y^{C[1]}_{s}  & = & Z_{s} + E^{[1]}_{s} \nonumber\\
Y^{C[m]}_{s}  & = & Z_{s} + Traj^{C[m]}_{s} + E^{[m]}_{s}. \nonumber
\end{eqnarray}
Implicitly, the fixed effects in this model are
\begin{eqnarray}
E(Z_{r}) = \alpha^{Rx} & = & E(Z_{s}) = \alpha^{C} \label{eq:EqualAlpha}\\
E(Traj^{Rx[m]}_{r}) = \mu^{Rx[m]}_{r} & \mathrm{and} & E(Traj^{C[m]}_{s}) = \mu^{C[m]}_{s}.
\end{eqnarray}

\textit{Change}, or change-from-baseline, is measured by subtracting the baseline (visit 1) measurement from the last (visit $ m $) measurement, so it does not have a random intercept component. \textit{Change} of patient $ i $ given treatment $ Trt $ is then
\begin{eqnarray*}
Y^{Trt[\mathrm{change}]}_{i} = \left[ Traj^{Trt[m]}_{i} \right] + \left[ E^{[m]}_{i} - E^{[1]}_{i} \right]
\end{eqnarray*}
so
\begin{eqnarray}
Var(Y^{Trt[\mathrm{change}]}_{i}) & = & Var( E^{[1]}_{i}) + Var( Traj^{Trt[m]}_{i}) + Var( E^{[m]}_{i}) \label{eq:VarChange}
\end{eqnarray}
which leads to
\begin{eqnarray}
Var(Z_{i}) + Cov(Z_{i} , Traj^{Trt[m]}_{i} )
& = &
\dfrac{Var(Y^{Trt[m]}_{i}) + Var(Y^{Trt[1]}_{i}) - Var(Y^{Trt[\mathrm{change}]}_{i})}{2}  \label{eq:VarZest}
\\
& = & Cov(Y^{Trt[m]}_{i}, Y^{Trt[1]}_{i}). \label{eq:CovVisits1m}
\end{eqnarray}
In the case where $ Z $ and $ Traj $ are independent, $ Cov(Z_{i} , Traj^{Trt[m]}_{i} ) = 0 $, we can estimate $ Var(Z_{i}) $ from (\ref{eq:VarZest}), and estimate $ Var(Traj^{[m]}_{i}) $ from (\ref{eq:Yafter}). In turn, we can estimate the variance of measurement error as
\begin{eqnarray}
Var(E^{[1]}_{i}) & = & Var(Y^{Trt[1]}_{i}) - Var(Z_{i}) \label{eq:VarEest}\\
& = & Var(E^{[m]}_{i}). \nonumber
\end{eqnarray}
Note that, if $ Z $ and $ Traj $ are in fact positively correlated (say), then estimates of $ Var(Z_{i}) $ and $ Var(Traj^{[m]}_{i}) $ would be somewhat higher than they ought to be, while the estimate for $ Var(E^{[1]}_{i}) = Var(E^{[m]}_{i}) $ would be somewhat lower than it ought to be.

\subsection{Providing input to the ETZ decomposition}
The ETZ decomposition (only) requires knowledge of the three quantities: $ Var(Y^{Trt[1]}_{i}) $, $ Var(Y^{Trt[m]}_{i}) $, and $ Var(Y^{Trt[\mathrm{change}]}_{i}) $.

If patient-level data is available, then after modeling data using Proc Mixed REPEATED (for example), $ Var(Y^{Trt[1]}_{i}) $ and $ Var(Y^{Trt[m]}_{i}) $ can be directly obtained as the $ [1,1] $ and the $ [m,m] $ elements of the $ \mathbf{R} $ matrix. Since $ Cov(Y^{Trt[m]}_{i}, Y^{Trt[1]}_{i}) $ is the $ [1,m] $ element of the $ \mathbf{R} $ matrix, one can then use the equality between (\ref{eq:VarZest}) and (\ref{eq:CovVisits1m}) to obtain $ Var(Y^{Trt[\mathrm{change}]}_{i}) $ as $ Var(Y^{Trt[1]}_{i}) + Var(Y^{Trt[m]}_{i}) - 2Cov(Y^{Trt[m]}_{i}, Y^{Trt[1]}_{i} )$. Obtaining the three variance estimates needed by ETZ from the same estimation process provides consistency.

If patient-level data is not available, then we obtain estimates of these three variances as best we can, from summary statistics reported for clinical trials, whatever form they may be in. For EXPEDITION3, we use the standard deviation (SD) of ADAS-Cog and ADCS-iADL at visit 1 and at visit $ m $ (Week 80) for both $ Rx $ and $ C $ provided in Table 2 of \cite{honig2018trial}. Note that, in this case, both visit 1 and visit $ m $ (Week 80) estimates are computed from \textit{raw scores}. While the estimate of $ Var(Y^{Trt[1]}_{i}) $ would be computed from raw scores, if the estimate of $ Var(Y^{Trt[m]}_{i}) $ reported for a clinical trial is computed from \textit{last observation carried forward} (LOCF) instead, one could use that. Also, as was the case for EXPEDITION3, typically standard error (SE) of least squares mean (LSmean) of \textit{Change} is reported instead of $ Var(Y^{Trt[\mathrm{change}]}_{i}) $. In such a situation, we convert the reported SE(\textit{Change}) to an estimate of $ Var(Y^{Trt[\mathrm{change}]}_{i}) $ using the reported sample sizes for $ Rx $ and $ C $ at visit 1 and visit $ m $, in a fashion similar to how one would using Hedge's formula in meta analysis.
\end{BlueText}

\section{Partition to control the incorrect decision rate}
\begin{figure}[H]
\centering
\includegraphics[width=0.8\linewidth]{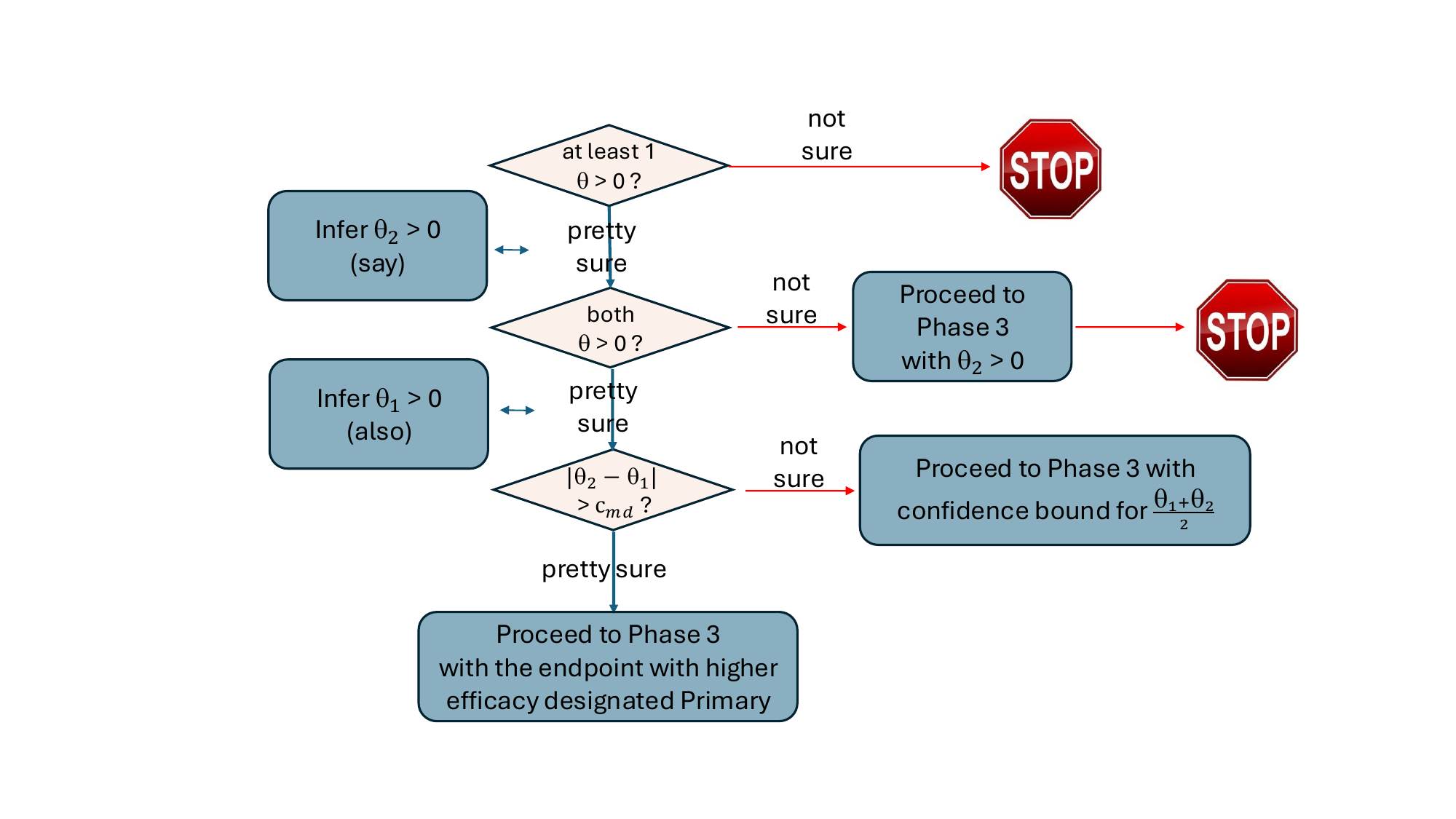}
\caption{This flowchart illustrates the decision path for a Phase 2 study with two endpoints, guiding (1) whether to transition to Phase 3, and (2) based on inference on the efficacy of the two endpoints, whether to designate one endpoint as primary or to combine both endpoints into a single composite endpoint.}
\label{fig:DecisionPathFlowChart}
\end{figure}
In the two-endpoint study setting, this article aims to address the following two questions:
\begin{enumerate}[leftmargin=1.4em]
\item \textbf{Confirmatory Transition Question:} Whether at least one of the endpoints shows positive efficacy; and
\item \textbf{Endpoint Designation Question:} If both endpoints show efficacy, which of the two should be designated as the primary endpoint, or whether a combination of the two should serve as the primary endpoint due to their comparable efficacy.
\end{enumerate}

\paragraph*{Testing equality nulls may not control Incorrect Decision rate}
In testing a single parameter $\theta$, one cannot assume controlling the Type I error rate for testing $H^=_{0}: \theta_{} = 0 $ automatically controls the directional error rate, because directional errors are not counted in the Type I error definition of testing an equality null. In everyday practice, danger of rejecting for wrong reasons when the hypotheses do not cover the entire parameter space is real. Section 7 of \cite{Liu&Tian&Hsu(2021)} shows, for a level-5\% log-rank test commonly used in survival analysis to test the null hypotheses that patients treated with $Rx$ and $C$ have equal survival probability within every time interval between events, the incorrect decision rate of deciding $Rx$ increases the median survival time relative to $C$ when in truth it does not exceed 15\%.

\paragraph*{The Partitioning Principle for hypothesis testing}
Partition the entire parameter space into null hypotheses where (potentially different) Type I error can occur, as well as (possibly) an alternative hypothesis where no Type I error can occur. By definition of \cite{Tukey(1953)} and \cite{Tukey(1994)}, Error Rate Familywise (often phrased as Familywise Error Rate, abbreviated as FWER, in current practice), testing each partitioned null hypothesis at level-$\alpha$ controls FWER (strongly), over the entire parameter space.

In contrast to decision-making by testing equality nulls, making decisions based on a confidence set controls the Incorrect Decision rate:

\begin{theorem}\label{thm:ConfidentInference}
If inferences are made with no point contained in a $ 100(1-\alpha)\% $ confidence set contradicting them, then one can be confident that all such inferences, no matter how many, will be simultaneously correct with probability $ (1-\alpha) $.
\end{theorem}
\begin{proof}
By definition, a $ 100(1-\alpha)\% $ confidence set contains the truth with probability $ (1-\alpha) $. Therefore making inferences (with statements such as parameters are greater or smaller than specific thresholds) that do not contradict a $ 100(1-\alpha)\% $ confidence set will not result in any incorrect inference more than $\alpha$ proportion of the time.
\end{proof}

\subsection{Partition the parameter space to construct confidence sets}\label{sec:PartitionConfidenceSet}
The so-called Neyman Construction of a confidence set (\cite{Neyman(1937)}) proceeds by testing for every possible parameter value \emph{in the entire parameter space} the null hypothesis that it is the true parameter. Since there is only one true parameter value, testing each null hypothesis at level-$\alpha$ controls the incorrect decision at $\alpha$. Collecting all the parameter values for which their corresponding nulls fail to be rejected results in a set $C$ that we call a ``confidence set'', a (random) set which contains the true parameter value with a probability of at least $1-\alpha$. A formal proof of this straightforward fact is given on pages 421-422 of \cite{CasellaBerger(2001)}.

Tukey’s and Dunnett’s methods are confidence sets in \textit{multi-dimensions} conveniently constructed by pivoting a multivariate-t statistic whose distribution depends neither on the unknown means of interest nor the nuisance variance-covariance parameters as follows. In the $p$-dimensional space, denote $(\theta_{1}, \ldots , \theta_{p})$ by $\btheta$ and its point estimate $({\hat{\theta}}_{1}, \ldots , {\hat{\theta}}_{p})$ by $\hat{\btheta}$. Suppose that the event $\{(\hat{\btheta}-\btheta) \in \Ba\}$ has a distribution that does not depend on any unknown parameter, with $\Ba$ scaled so that $P\{(\hat{\btheta}-\btheta) \in \Ba\} = 1-\alpha$. Then the following 2-step process results in a level-(1-$\alpha$) confidence set for $\btheta$:
\begin{description}[leftmargin=1.2em]
\item[Reflection] Reflect $\Ba$ through the origin $\mathbf{0}$ to obtain $-\Ba$;
\item[Translation] Shift the origin $\mathbf{0}$ in $-\Ba$ to be at the point estimate $\hat{\btheta}$.
\end{description}
However, Tukey's and Dunnett's confidence sets do not target comparisons of specific interest in different parts of the parameter space. Amazingly, in the 1970s, \cite{Takeuchi(1973)} developed sophisticated confidence sets targeting the determination of signs of $\theta_{1}, \ldots, \theta_{p}$. But since it was published in the Japanese language, his impressive work was largely unknown until re-discovered and reported at the 2007 MCP (Multiple Comparison Procedures) conference in Vienna, and honored at the 2009 MCP conference in Tokyo. Happily, we can now read this work in English in \cite{Takeuchi(2010)}.

Our Pivoting-Partitioning Confidence set construction below first uses Neyman Construction to partition the multi-dimensional parameter space into regions with different comparisons of interest in each, then uses pivoting to conveniently construct confidence subset in each region \textit{directed} toward those comparisons, which are then coalesced into a confidence set for $\btheta$ in the entire parameter space.

\begin{theorem}[Pivoting-Partitioning Confidence Set Construction]\label{lemma.reflect}
Partition the entire parameter space $\Theta$ into sub-spaces $\Theta_1 , \ldots, \Theta_M$. If each $A_1, \ldots, A_M$ is scaled so that $P\{(\hat{\btheta}-\btheta) \in A_{m}\} = 1-\alpha$, then a level {\rm 100}$(1- \alpha)$\% confidence region for $\hat{\btheta}$ is
\[
C({\hat{\theta}}_{1}, \ldots , {\hat{\theta}}_{p})
= \bigcup_{m=1}^M \left( \{-\btheta + 
{\hat{\btheta}} : -\btheta + 
{\hat{\btheta}} \in A_m\} \cap \Theta_m \right).
\]
\end{theorem}

\begin{proof}
Denote by $\btheta^{0}$ a candidate true parameter value. By Neyman Construction, a level $100(1 - \alpha)\%$ confidence set for $\btheta$ is
\begin{align*}
    C({\hat{\btheta}}) & =  
\{\btheta^{0} \in \Theta : H_0: \btheta = \btheta^{0} \mbox{ is not ejected }\}\\
& =  \bigcup_{m=1}^M \{\btheta^0 \in \Theta_m :  {\hat{\btheta}}-\btheta^{0}\in A_m \}\\
& =  \bigcup_{m=1}^M \left( \{\btheta
\in -A_m + \hat{\btheta}\} \cap \Theta_m \right).
\end{align*}
\end{proof}

As an example of how Pivoting-Partition confidence set construction has proven useful, suppose $\theta_1$ and $\theta_2$ are the primary and secondary endpoints respectively, and values no greater than zero indicate a lack of efficacy. \cite{Hsu&Berger(1999)} partitioned the parameter space as $\Theta_1 = \{\theta_1 \le 0\}$, $\Theta_2 = \{\theta_2 \le 0 \mbox{ but } \theta_1 > 0 \}$ and $\Theta_3 = \{\theta_1 > 0 \mbox{ and } \theta_2 > 0\}$. Since they are disjoint, no multiplicity adjustment is needed in testing $H_{0m}:\btheta \in \Theta_m$. As rejecting $H_{02}$ indicates efficacy in the secondary endpoint only if $H_{01}$ is rejected as well, this leads to testing $H_{01}$ first and only if it is rejected then test $H_{0s}$. As another example, \cite{Stefansson&Kim&Hsu(1988)} used Pivoting-Partitioning to construct a confidence set for the \textit{step-down version} of Dunnett's method for comparing new treatments with a control.

\paragraph{Insight}
Partition testing of hypotheses and confidence set construction are \emph{conditional} inferences, conditioning on \emph{relevant subsets}, using observed data to eliminate subsets of the partitioned parameter space. Basis of the Partitioning Principle was Multiple Comparisons with the Best (MCB) as described in Chapter 4 of \cite{Hsu(1996)}, which Jason Hsu developed by connecting the idea of \emph{conditional confidence} in Ranking and Selection in Section 4 of \cite{Kiefer(1977)} with John W. Tukey's Multiple Comparisons framework. \textit{Relevant subset} Neyman Construction confidence sets are unlikely to have issues such as the \emph{marginalization paradox} that afflict Fiducial/Structural confidence sets constructed by conditioning on \emph{ancillary} statistics (such as \emph{residuals}). Further, whereas Kiefer changed the \emph{conditional confidence coefficient} (to higher or lower than the nominal confidence level) conditionally while keeping the shape of the confidence set constant, \emph{directed} Partitioning confidence sets \emph{shape-shift} to keep the conditional confidence level as constant as possible.

In order to answer the \textbf{Confirmatory Transition Question} and \textbf{Endpoint Destination Question} above, the two-dimension parameter space $\Theta \subseteq \mathbb{R}^2$ for $\theta_1$ and $\theta_2$ should be further partitioned into:
\begin{eqnarray}
\Theta^{\le \le}: \theta_{1} \le 0 & \mbox{and} & \theta_{2} \le 0 \label{ParRegn: NegNegQuadrant}\\
\Theta^{  > \le}: \theta_{2} \le 0 & (\mbox {but} & \theta_{1} > 0 ) \label{ParRegn: PosNegQuadrant}\\
\Theta^{\le   >}: \theta_{1} \le 0 & (\mbox{but} & \theta_{2} > 0 )\label{ParRegn: NegPosQuadrant}\\
\Theta^{> >\nabla}: \theta_{1} > 0 & \mbox{and} & \theta_{2}-\theta_{1} > c_{md} \label{ParRegn: PosPosQuadrant-UpperCone}\\
\Theta^{> >band}: \theta_{2}-\theta_{1} \le c_{md} & \mbox{and} & \theta_{2}-\theta_{1} \ge -c_{md} \label{ParRegn: PosPosQuadrant-Band}\\
\Theta^{> >\triangle}: \theta_{2}-\theta_{1} < -c_{md} & \mbox{and} & \theta_{2} >0\label{ParRegn: PosPosQuadrant-BottomCone}
\end{eqnarray}
where the $ c_{md} $ is a clinically meaningful difference predefined by clinicians.

\begin{figure}[H]
\centering
\includegraphics[width=1\linewidth]{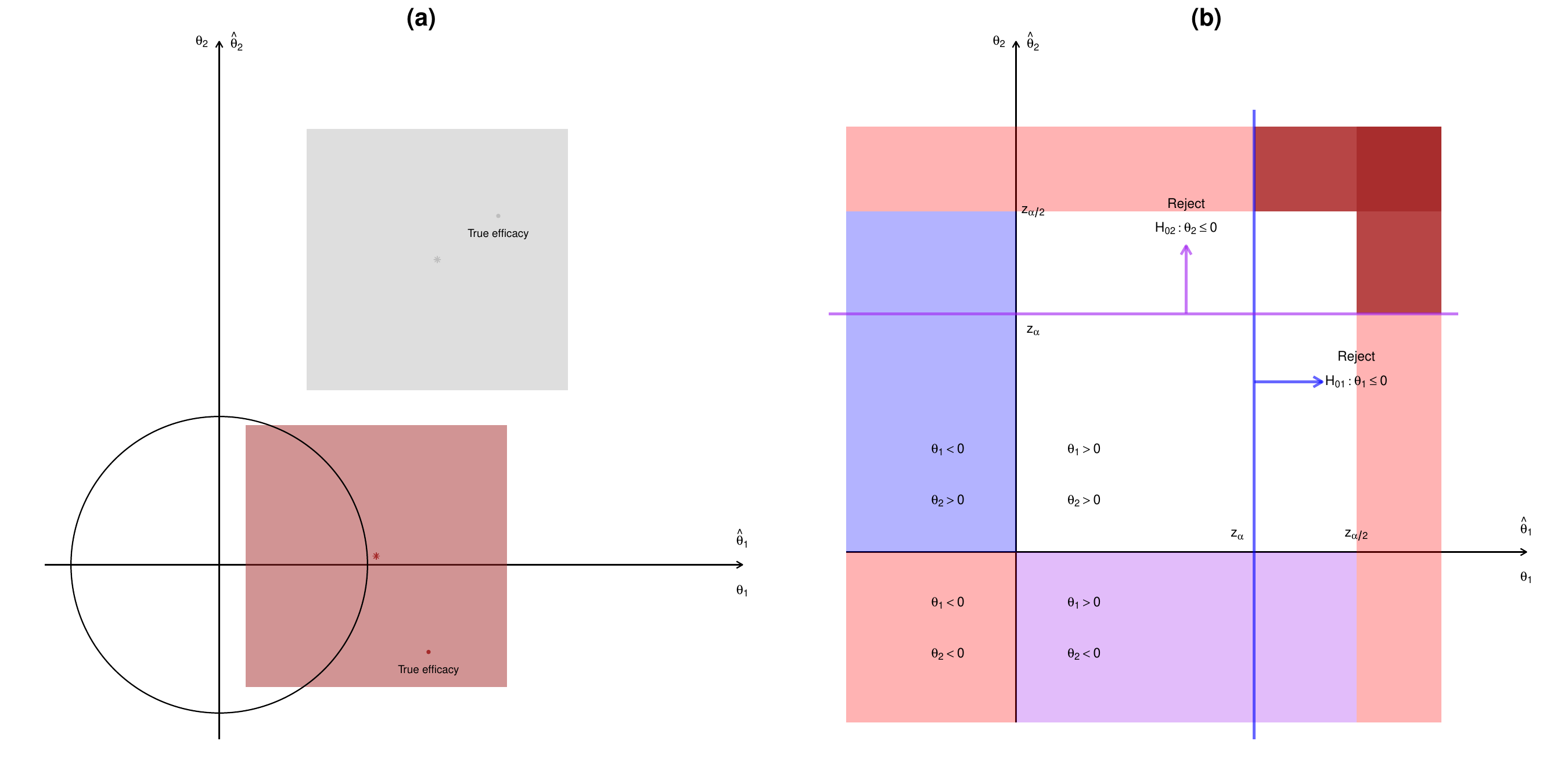}
\caption{Correct and incorrect decision‐making involving two efficacy parameters $\theta_1$ and $\theta_2$ (that are efficacious when positive). Plot \textbf{(a)} shows two true parameter points (red, gray) with their confidence rectangles. The $F$‐test rejects $H_0\colon\theta_1=\theta_2=0$ when $(\hat\theta_1,\hat\theta_2)$ lies outside the rejection circle. If the red point is the truth, the F-test would reject when $(\hat\theta_1,\hat\theta_2)$ is inside the gray rectangle perceiving incorrectly $\theta_2>0$ (leading to power calculation that includes rejecting for wrong reasons). In contrast, making decisions based on confidence sets yields confident two‐sided Correct and Useful inference by Theorem \ref{thm:ConfidentInference}. For example, the red confidence rectangle correctly infers $\theta_1>0$ only. Plot \textbf{(b)} shows typical one‐sided rejection regions for the partitioned quadrants $\Theta^{\le\le}\!:\,\theta_1\le0,\theta_2\le0$; $\Theta^{>\le}\!:\,\theta_1>0,\theta_2\le0$; and $\Theta^{\le>}\!:\,\theta_1\le0,\theta_2>0$, with rejection of $\Theta^{\le\le}$ requiring multiplicity adjustment with bounds $b_{\rho_1},b_{\rho_2}\in[z_\alpha,z_{\alpha/2}]$, while rejections of the single‐efficacy quadrants do not.}
\label{fig:CombinedEfficacy}
\end{figure}

(\ref{ParRegn: NegNegQuadrant}), (\ref{ParRegn: PosNegQuadrant}) and (\ref{ParRegn: NegPosQuadrant}) correspond to the plot (b) of Figure \ref{fig:CombinedEfficacy} while (\ref{ParRegn: PosPosQuadrant-UpperCone}), (\ref{ParRegn: PosPosQuadrant-Band}) and (\ref{ParRegn: PosPosQuadrant-BottomCone}) can be further partitioned in Figure \ref{fig:Pos-posQuadrant}.

(\ref{ParRegn: NegNegQuadrant}), (\ref{ParRegn: PosNegQuadrant}) and (\ref{ParRegn: NegPosQuadrant}) are partitioned to answer the \textbf{Confirmatory Transition Question}. A Phase 2 study can transition to a Phase 3 study when the Negative-negative quadrant (\ref{ParRegn: NegNegQuadrant}) and at least one of the Positive-negative quadrant (\ref{ParRegn: PosNegQuadrant}) and the Negative-positive quadrant (\ref{ParRegn: NegPosQuadrant}) are not covered by confidence set. The contents in the brackets of (\ref{ParRegn: PosNegQuadrant}) and (\ref{ParRegn: NegPosQuadrant}) can be omitted because whether an endpoint shows efficacy is an one-sided problem.

\paragraph*{The concept of \emph{directed} confidence sets}
This concept of \textit{directed} confidence set in \cite{Hsu&Berger(1999)} is to shape the confidence set within each of partitioned region of the parameter space toward the inference desired for that region.

In Bioequivalence, the desired inference is $-\delta \le \theta \le \delta$ where $\delta$ is the equivalence margin. So the Neyman Construction of \cite{Hsu&Hwang&Liu&Ruberg(1994)} tested for each $\theta^0 > 0$ the simple null $H_0: \theta = \theta^0$ against the simple alternative $H_a: \theta = 0$ at level-$\alpha$, while for each $\theta^0 < 0$ they also tested the simple null $H_0: \theta = \theta^0$ against the alternative $H_a: \theta = 0$ at level-$\alpha$. That each of these two sets of 1-sided tests in opposite directions is most powerful according to the Neyman-Pearson lemma makes the resulting Bioequivalence confidence interval have the smallest expected length among all level-($1-\alpha$) confidence intervals if $\theta = 0 $ in truth.

Instead of desiring ``equivalence'', to decide which endpoint should be designated Primary and the other Secondary in our Confirmability setting, the desired inference is \textit{meaningful difference}. The directed Neyman Construction confidence set for that is given in Section 4.1 of \cite{Hayter&Hsu(1994)}.

Deciding whether to proceed to a confirmatory study, is a complex decision-making process involving multiple parameters.

Our view is that there are two possible strategies for choosing two endpoints:
\begin{enumerate}[leftmargin=1.4em]
\item[$ \upuparrows $] If a compound's efficacy is comparable in magnitude and variability in both endpoints (as in the Table \ref{table:EXPEDITION3endpointsCombine}), then combining the two endpoints into a single endpoint is reasonable both medically and as a strategy for gaining marketing approval;
\item[$ \perp $] if efficacy in the two endpoints are rather different, substantially higher in one than the other, then a sponsor would want to designate the stronger one to be the primary endpoint and the weaker one to be the secondary endpoint.
\end{enumerate}

\begin{table}[H]
\centering
\begin{tabular}{|c|c|c|}
\hline
Endpoint & Estimated efficacy & 95\% Confidence Interval \\
\hline
ADAS-Cog14 & $-0.80$ & ($-1.73 , 0.14$) \\
\hline
ADCS-iADL & $1.00$ & ($0.17, 1.83$) \\
\hline
\end{tabular}
\caption{Solanezumab's efficacy in ADAS-Cog14 and ADCS-iADL are comparable in magnitude and variability in EXPEDITION3}
\label{table:EXPEDITION3endpointsCombine}
\end{table}

In July 2024, donanemab (Kisunla${\textsuperscript{\texttrademark}}$) was approved by the FDA for treating patients with early symptomatic Alzheimer's disease. Its primary outcome measure iADRS (integrated Alzheimer’s Disease Rating Scale) combines ADAS-Cog13 and ADCS-iADL. While a higher ADCS-iADL is better, a lower ADAS-Cog is better. So iADRS is calculated by subtracting ADAS-Cog13 from ADCS-iADL, with a higher iADRS being better (less impaired). To anchor ADAS-Cog at zero, a constant 85 is added for ADAS-Cog13 (90 for ADAS-Cog14), see \cite{WesselsEtAl(2015)}.

As one does not know whether the truth is $ \upuparrows $ or $ \perp $, we design the acceptance/rejection regions to achieve a ``directed'' confidence set.

\subsection{Meaningful difference confidence sets construction}
\begin{figure}[htbp]
\centering
\includegraphics[width=1\linewidth]{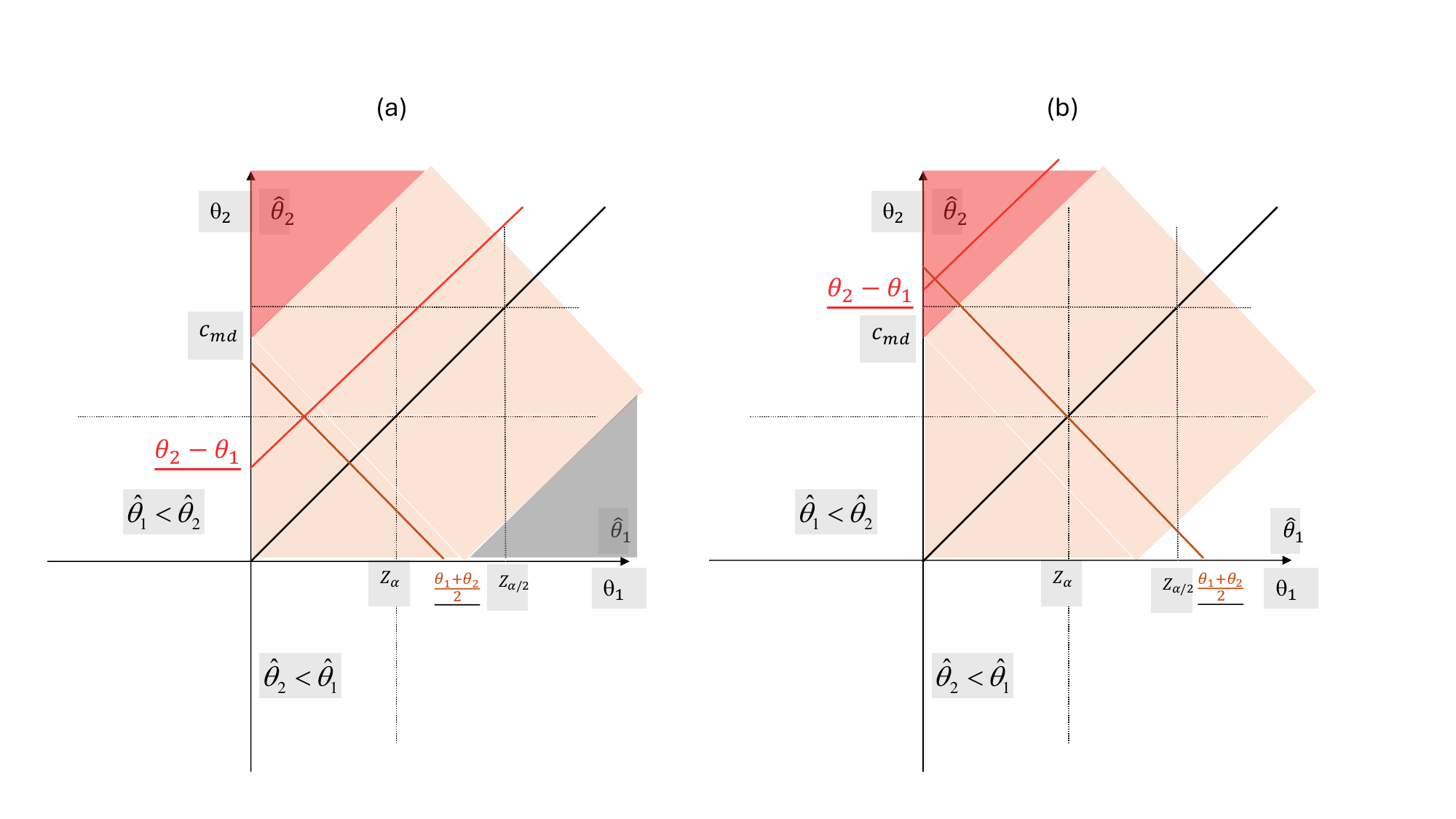}
\caption{This figure presents two situations in the positive-positive quadrant, partitioned into three regions: the scarlet top cone $\Theta^{> >\nabla}$, the grey bottom cone $\Theta^{> >\triangle}$, and the beige middle band region $\Theta^{> >band}$. In the parameter space, if one can identify either $\theta \in \Theta^{> >\nabla}$ or $\theta \in \Theta^{> >\triangle}$, then endpoint 2 or 1 should be designated as the primary endpoint, respectively. Otherwise, one possible strategy is to combine the two endpoints into a single composite endpoint. For example, while Plot (a) suggests the two endpoints be combined into a single endpoint, Plot (b) indicates endpoint 2 should be designated as the primary endpoint.}
\label{fig:Pos-posQuadrant}
\end{figure}

To answer the \textbf{Endpoint Designation Question}, the Positive-positive quadrant is partitioned into (\ref{ParRegn: PosPosQuadrant-UpperCone}), (\ref{ParRegn: PosPosQuadrant-Band}), and (\ref{ParRegn: PosPosQuadrant-BottomCone}), respectively, as shown in Figure \ref{fig:Pos-posQuadrant}.
Confidence set decision-making for answering this question has two layers as was shown in Figure \ref{fig:DecisionPathFlowChart}.

The first possibility is one endpoint shows significantly higher efficacy than the other, as indicated by all candidate values in either the top cone or the bottom cone in Figure \ref{fig:Pos-posQuadrant} are eliminated from the confidence set, so which endpoint to designate as primary is clear. The second possibility is difference in efficacy between the two endpoints is not clinically meaningful, in which case a lower confidence bound of the combined efficacy (averaged over the endpoints) is computed.

Confidence construction for the top and bottom cones inspired by Neyman construction and section 4.1 in \cite{Hayter&Hsu(1994)} is as follows.

\paragraph*{Intuition for meaningful difference confidence construction}
As discussed in the insight for Theorem \ref{thm:ConfidentInference}, Neyman's confidence set construction permits each partitioned region to adopt its own most informative acceptance region shape.

For each of the partitioned top and bottom cone regions, the desired inference is 1-sided, aiming to exclude parameter values lying in the opposite partitioned cone. Thus, the confidence set for the cone-shaped partitioned regions should be directed \emph{away from} the origin, as opposed to the bioequivalence confidence set described in Section \ref{sec:PartitionConfidenceSet} being directed \emph{toward} the origin.

\begin{figure}[H]
\centering
\includegraphics[width=0.8\linewidth]{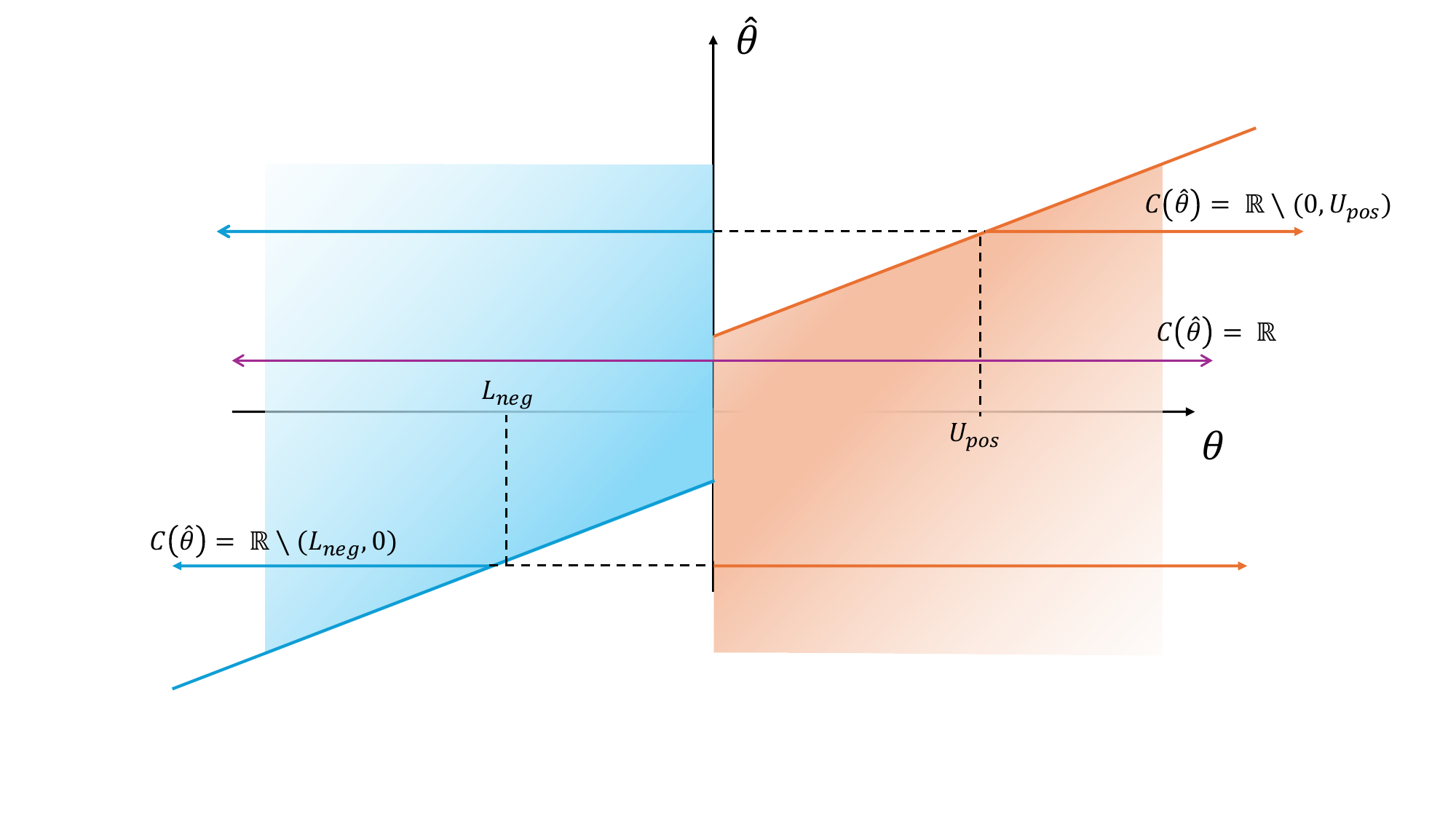}
\caption{This plot explains why one-sided acceptance regions are ineffective. The parameter space is partitioned into $\Theta^- = \{\theta:\theta \le 0\}$ and $\Theta^+ = \{\theta:\theta > 0\}$, and two one-sided acceptance regions in opposite directions, represented by regions of the two different colors, are considered for the partitioned $\Theta^-$ and $\Theta^+$. Regardless of $\htheta$, the Neyman construction confidence set cannot entirely eliminate either $\Theta^-$ or $\Theta^+$.}
\label{fig:OneSidedAllowancePlot}
\end{figure}

Surprisingly, for 1-sided meaningful difference inference, it is necessary to construct a confidence set using a 2-sided acceptance region. That is because, as shown in Figure \ref{fig:OneSidedAllowancePlot}, the confidence set resulting from an 1-sided acceptance region will always contain values from both cone-shaped partitioned regions, unable to \textbf{usefully} infer that one endpoint has meaningfully higher efficacy than the other. Only with a 2-sided acceptance region can the constructed confidence set conclude that a specific endpoint should be designated as the primary one.

We show Neyman construction of a confidence set for the top cone as an example.
\paragraph*{Example}
For each candidate parameter value in the top (scarlet) cone, a \emph{lower} bound of the acceptance region testing it is needed, so that if the true parameter value is in the lower (grey) cone, the entire top cone is eliminated. For that purpose, we can use the same fixed lower bound $-\sigma_{diff}z_{\alpha/2}$ as one would in traditional 2-sided testing, without tailoring it to the specific candidate parameter value.

However, when $\hat{\theta}_{diff}$ is able to eliminate the entire lower cone, we would want the confidence bound to be as positively away from the origin as possible. For that purpose, we let the upper bound of the acceptance region testing each candidate value to increase with it, so the asymmetric acceptance region becomes
\begin{eqnarray}
A(\theta_{diff}) & = & \{\hat{\theta}_{diff}: - \sigma_{diff}z_{\alpha/2} \leq \hat{\theta}_{diff} \leq \overline{e}_{\alpha,|\theta_{diff}|}\}\label{eq: WeakPartitionAcptRgn},
\end{eqnarray}
satisfying
\begin{equation}
P( - \sigma_{diff}z_{\alpha/2}\leq \hat{\theta}_{diff}\leq \overline{e}_{\alpha,|\theta_{diff}|}) = 1 - \alpha,\label{eq: AccRegionProbForm-weak}
\end{equation}
where $\hat{\theta}_{diff} \sim N(\theta_{diff},\sigma_{diff}^2)$ with $\overline{e}_{\alpha,|\theta_{diff}|}$ increasing monotonically from $\sigma_{diff}z_{\alpha/2}$ to $\theta_{diff}+\sigma_{diff}z_{\alpha}$ as $\theta_{diff}$ increases from 0 to $+\infty$. The resulting directed confidence set for the top cone is a one-sided confidence interval $\theta_{diff} \in [\hat{\theta}_{diff}-d_{\alpha,|\hat{\theta}_{diff}|},+\infty)$ when $\hat{\theta}_{diff}>\sigma_{diff}z_{\alpha/2}$ with $d_{\alpha,|\hat{\theta}_{diff}|}$ being the solution to
\begin{equation}
\Phi\!\left(\frac{d_{\alpha,|\hat{\theta}_{diff}|}}{\sigma_{diff}}\right) - \Phi\!\left(\frac{( - \sigma_{diff}z_{\alpha/2}) - (\hat{\theta}_{diff} - d_{\alpha,|\hat{\theta}_{diff}|})}{\sigma_{diff}}\right) = 1 - \alpha,
\end{equation}
but is a non-informative $\theta_{diff} \in (-\infty,+\infty)$ when $|\hat{\theta}_{diff}|\le \sigma_{diff}z_{\alpha/2}$ (corresponding to the situation that traditional 2-sided testing fails to reject $H_0: \theta_{diff} = 0$).

\begin{figure}[htbp]
\centering
\includegraphics[width=0.5\linewidth]{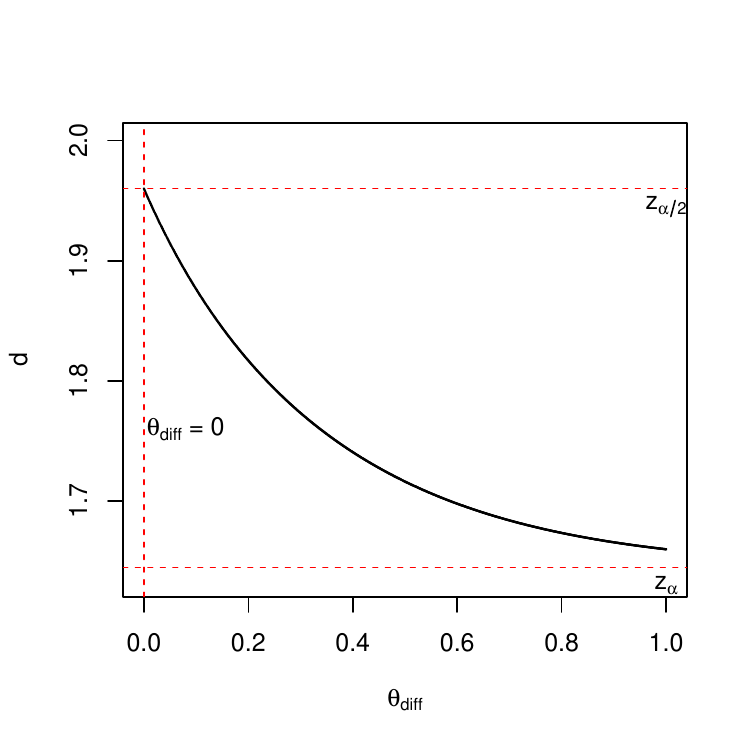}
\caption{The allowance $d$ decreases from $z_{\alpha}$ approximately to $z_{\alpha/2}$ as $ \theta_{diff}$ increases from $0^+$ to $+\infty$.}
\label{fig:dThetaTrend}
\end{figure}
A directed confidence set for the bottom (grey) cone can be constructed analogously.

Consider that the two endpoints are ADCS-iADL and ADAS-Cog14. The difference in efficacy can be denoted as $\theta_{diff} = \theta_{iADL} - \theta_{Cog}$, where $\theta_{iADL}$ is the scaled efficacy for ADCS-iADL and $\theta_{Cog}$ is the opposite of the scaled efficacy for ADAS-Cog14. Both $\theta_{iADL}$ and $\theta_{Cog}$ are scaled such that higher values indicate better efficacy.

Using the estimates from Table~\ref{table:EXPEDITION3endpointsCombine} for solanezumab as an example, and starting with the assumption of independence as a baseline, we have the estimated difference in efficacy $\hat{\theta}_{diff} = 0.2$, with an estimated standard deviation of $\hat{\sigma}_{diff} = 0.98$.

Since $\hat{\theta}_{diff} < \hat{\sigma}_{diff} z_{\alpha/2}$, the confidence interval for $\theta_{diff}$ includes both cone-shaped partitioned regions. Thus, it is reasonable to consider combining the two endpoints, as is done with donanemab.

Define the average efficacy as $\theta_{avg} = \frac{\theta_{iADL} + \theta_{Cog}}{2}$. The estimated average efficacy is $\hat{\theta}_{avg} = 0.9$, with an estimated standard deviation of $\hat{\sigma}_{avg} = 0.49$. The lower bound of the confidence interval is 0.09, which is positive, supporting the designation of the combined efficacy as a primary endpoint.

\section{Phase 2 to 3 Transition Decision-making Process}\label{sec:DMProcess}
Consider designing a confirmatory study to meet a high prespecified probability of ``success''. Suppose the statistical analysis plan has a 5\% chance of making an incorrect inference. Then for any design to claim that it has a higher than 95\% chance, 97\% ``power'' say, of leading to a ``success'' is unreasonable, because in that case the definition of ``success'' evidently includes at least a 2\% chance of erroneous or false ``success''.

That the power is set so high is perhaps due to experience with high Phase 2 to Phase 3 transition failure rate, as observed in \cite{HayEtAl(2014)} for example. Our understanding is often there is some off-the-books ``discounting'' of the calculated \textit{power}, to be conservative. With the ETZ modeling principle able to take all sources of variability into account, instead of informal discounting, this article offers a direct calculation of success confidence, which is the probability of true success in transitioning from a feeder study to a confirmatory study (in analogy to a confidence interval guaranteeing coverage probability).

\subsection{Replacing \texorpdfstring {\(P\{\mathit{correct\ and\ useful\ inference}\}\)}{P\{correct and useful inference\}}}
The concept of ``power'' is meant for 2-action problems such as testing a simple null hypothesis against a simple alternative. For problems with multiple parameters, such calculation of ``power'' is legacy practice from B.C. (before computers) when comparing $k$ means typically used the \textit{F}-test for testing $H_0: \mu_1 = \ldots = \mu_k$ in an ANOVA setting (e.g., \cite{Scheffe(1959)} Section 3.3). However, as \cite{Hsu(1989)power} pointed out, included in the power calculation of an omnibus test is the probability of \emph{rejecting for wrong reasons}. Suppose, in comparing high, medium, and low doses of a psychiatric drug against a placebo, the truth is the medium dose is effective but high and low doses are not. Power calculation of the \textit{F}-test (e.g., by Proc Power and Proc GLMpower in SAS) includes the probability that it rejects for the wrong reason that data indicates incorrectly high or low dose is effective but the medium dose is not, for example. With modern computing power, we believe this legacy practice should be ditched in view of the following.

\paragraph*{The Correct and Useful inference principle}
Statistical inferences are truly useful only if they are correct.

To follow this principle, \cite{Hsu(1989)power} and Appendix C of \cite{Hsu(1996)} defined the result of a study to be correct and useful if the following two events simultaneously occur:
\begin{description}[leftmargin=1.2em]
\item[Correct =] \{no new treatment worse (better) than the control will be incorrectly inferred to be better (worse) than the control\}
\item[Useful =] \{all treatments sufficiently better (worse) than the control will be inferred to be better (worse) than the control\}
\end{description}
Clearly, if $ \inf P\{\mathrm{Correct}\} = 95\%$, and $ P\{\mathrm{Useful}\} < 100\% $, then $ P\{ \{\mathrm{Correct}\} \cap \{\mathrm{Useful}\} \} = P\{\mathrm{Correct~and~Useful~inference}\} < 95\% $ which is one aspect of the \textit{correct} and \textit{useful} inference concept more reasonable than that of ``power''.

To make the event $ \{\mathrm{Correct~and~Useful~inference}\} $ occur with sufficiently high probability requires replacing omnibus testing by simultaneous inference methods such as Tukey's and Dunnett's which are capable of inferring the magnitude and direction of individual treatment efficacy. These methods only became computationally feasible for the General Linear Model from the development of the Factor Analytic algorithm by \cite{Hsu(1992)GLM}. The ProbMC function in SAS, which implements the FA algorithm, can be used to calculate the sample size needed to achieve the desired $  P\{\mathrm{Correct~and~Useful~inference}\} $.

As illustrated in Figure \ref{fig:CombinedEfficacy}, operationally, if the clinically meaningful threshold for a (component) parameter $ \theta_i $ is $ \delta_i $, then one infers $ \theta_i > \delta_i $ only if the confidence interval for $ \theta_i $ (given by the confidence set) is entirely larger than $ \delta_i $ (similarly for inferring $ \theta_i < \delta_i $). With confidence intervals covering their true values guaranteeing \textit{correctness}, one can design a study ensuring the event $ \{\mathrm{Correct~and~Useful~inference}\} $ occurs with desired probability by having sufficient sample size so that the confidence intervals will be sufficiently tight.

\subsection{Acknowledging uncertainty in true efficacy}
To decide whether it is a correct decision to proceed from a feeder study to a confirmatory study, current industry practice seems to be to take estimates from the feeder study (and other sources) as the truth, and compute the sample size needed to adequately ``power'' the confirmatory study, using
\begin{equation}
n = 2(z_{\alpha/2} + z_\beta)^2 (\sigma/\Delta)^2 \label{eq:SampleSize}
\end{equation}
where $ \alpha $ is the probability of type I error, $ \beta $ is the probability of type II error, $ \Delta $ is the true separation between $ Rx $ and $ C $ at the milestone visit of the study (visit $ m $), $ \sigma $ is the true standard deviation of the outcome, $ z_\gamma $ denotes the upper $ \gamma $ quantile of the standard Normal distribution.

For example, page 260 of the published protocol of EXPEDITION3 states
\begin{quote}
``Based on data from completed Phase 3 solanezumab Studies LZAM and LZAN $ \dots $ in mild AD patients with evidence of amyloid pathology $ \ldots $ for ADAS-Cog14, we assume a treatment difference of approximately 2 points at 18 months with a standard deviation of 10. For the ADCS-iADL, we assume a treatment difference of 1.5 points at 18 months with a standard deviation of 10. $ \ldots $ 1050 randomized patients per arm, or 2100 randomized patients total $ \ldots $, will have approximately 97\% power for the ADAS-Cog14 comparison and approximately 84\% power for the ADCS-iADL comparison to detect a significant treatment difference at 18 months using a 2-sided significance level of 0.05.''
\end{quote}

\begin{figure}[htbp]
\centering
\includegraphics[width=0.9\linewidth]{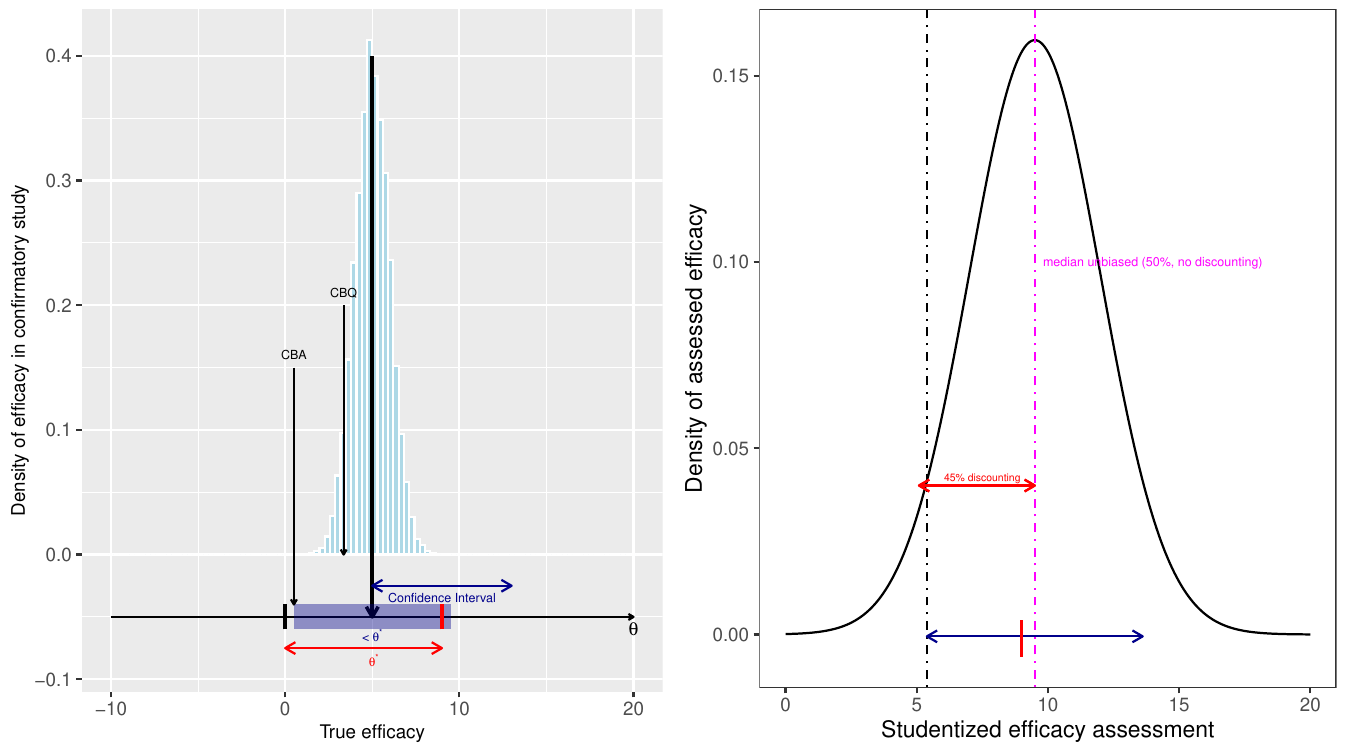}
\caption{The plot on the left panel shows Confidently Bounded Quantile (CBQ) versus Confidently Bounded Allowance (CBA). The horizontal axis represents efficacy with the \textbf{short vertical red segment} being the true efficacy. The double-arrow horizontal bar represents what a 95\% confidence interval from the feeder (Phase 2) study centered around the true value might look like. The semi-transparent dark blue area visualizes a possible Tukey's CBA with a confidence interval rather off-centered but covers the true value (so the inference will be correct) and is sufficiently narrow to not cover zero (so the inference will be useful). On the other hand, our CBQ takes a conservative limit of the \textbf{blue} double-arrow confidence interval as the true efficacy, and simulates a confirmatory study (Phase 3) 10000 times, resulting in the light blue histogram. CBQ is then the quantile of this histogram that ensures $ P\{\mathrm{True~ success}\} \ge \gamma$. The plot on the right panel shows Phase 2 discounting. A 95\% lower confidence limit of efficacy from Phase 2 data corresponds to a Phase 2 discounting of 45\%, relative to a 50/50 (median unbiased) point estimate. (A 50\% confidence limit has a 50/50 chance of being higher/lower than the true value.)}
\label{fig:Combinedplot}
\end{figure}

The concept of Correct and Useful inference is a principle applicable to complex decision-making, with CBA's \textit{coverage} being a form of \textit{correctness} and \textit{narrowness} being a form of \textit{usefulness} in the simple setting of inference on a single parameter. In the Confirmability setting, we define ``\emph{true} success'' as, at the end of the confirmatory study, a truly efficacious compound is correctly inferred to be efficacious. True success requires the occurrence of the two events
\begin{description}[leftmargin=1.2em]
\item[Transition] Estimated efficacy from a Phase 2 feeder study conservatively justifies transitioning,
\item[Confirmation] The Phase 3 study successfully confirms efficacy,
\end{description}
the probabilities of which depend not only on true efficacy but also on variabilities.

Our Confidently Bounded Quantile (CBQ) follows the correct and useful inference principle. CBQ is a decision-making process that designs a study with adequate sample size and control of variabilities to ensure a pre-specified success confidence $ P\{\mathrm{True~ success}\} \ge \gamma$ (e.g., 85\%).

Replacing off-the-books discounting of \textit{power}, CBQ is a formal 2-step conditional discounting process. The first discounting is, instead of a point estimate of true efficacy, a Confident Efficacy is estimated from Phase 2 data. Then, following the Conditionality Principle which states statistical inference should be based on the experiment actually performed (\cite{Berger(1985)}, page 31), \emph{conditional on this Confident Efficacy}, the second discounting calculates a \textit{Confidently Bounded Quantile} for the Phase 3 outcome taking confirmatory study variability into account.

To be precise, visualize making transitioning decisions over infinitely many Phase 2 to potential Phase 3 studies. Define
\begin{itemize}[leftmargin=1.4em]
\item \textbf{event $ A $} as the event that Phase 2 study transitions to Phase 3 study correctly, and
\item \textbf{event $ B $} as the event that Phase 3 study shows efficacy successfully.
\end{itemize}
$P(A\cap B)$, the overall probability of making a correct decision, equals $P(A)P(B \mid A),$ where $P(B\mid A)$ is the probability that the Phase 3 study shows efficacy conditional on the confidence interval for true efficacy from the Phase 2 covering the true value. Instead of using the Bonferroni inequality as in CBA, we calculate the multivariate probability precisely using modern computing facilities, interfacing with users in the form of an interactive app.

CBQ calculates a Confident Efficacy by taking a conservative confidence limit of true efficacy, then uses it to confidently estimate a Phase 3 outcome by replicating the Phase 3 study with a fixed sample size many times. To achieve confidence, two discountings are employed for distinct purposes:
\begin{itemize}[leftmargin=1.4em]
\item \textbf{Phase 2 discounting} confidently assesses efficacy from the feeder study, accounting for variability between Phase 2 studies.
\item \textbf{Phase 3 discounting} confidently predicts potential outcome of the Phase 3 study, accounting for variability replicating a Phase 3 study.
\end{itemize}
Each discounting is relative to a 50/50 (median unbiased) point estimate. As illustrated in Figure \ref{fig:Combinedplot}, a Phase 2 \textit{discount} of $d_{Phase2}$ controls $P(A)$ at $P(A) = d_{Phase2}+ 50\%$. For example, if the Phase 2 discounting is $ d_{Phase2} = 45\% $, then Phase 2 discounting (conservatively) calculates true efficacy as its lower 45\%+50\% = 95\% confidence limit from the Phase 2 data.

A Phase 3 \textit{discount} of $d_{Phase3 \mid Phase2}$ controls $P(B \mid A)$ at $P(B \mid A) = d_{Phase3 \mid Phase2}+50\%$. So, with $ P\{A \cap B\} = P(A) P(B \mid A) = (d_{Phase2}+ 50\%)( d_{Phase3 \mid Phase2}+50\%)$, given a (fixed) desired success confidence $ P\{\mathrm{True~success}\} = \gamma $, setting $d_{Phase2}$ determines $d_{Phase3 \mid Phase2}$, and setting $d_{Phase3 \mid Phase2}$ determines $d_{Phase2}$. For example, if $ \gamma = 80\% $ and one sets $d_{Phase2} = 45\%$, then $d_{Phase3 \mid Phase2} = 34.21\%$, so that $ P\{A \cap B\} = (45\%+50\%)(34.21\%+50\%) = 80\% $.

\subsection{Assessing EXPEDITION3 as a feeder example using the App}
Our Confirmability App is accessible at https://xwukstatsci.shinyapps.io/confirmability/. Its \textit{Transition Decision-Making Process Plots} tab calculates whether the confident quantile exceeds the desired threshold. We illustrate this using summary statistics from the EXPEDITION3 study described in Section \ref{sec:EXPEDITION3}. For the EXPEDITION3 study, with $ Rx $ being solanezumab and $ C $ being a placebo, Figure 2 from \cite{honig2018trial} shows the estimated profile for Alzheimer Disease Cooperative Study-Instrumental Activities of Daily Living (ADCS-iADL) is approximately linear, while the estimated profile for ADAS-Cog14 shows some non-linearity. The confidence interval for the difference of change-from-baseline between $ Rx $ and $ C $ at 80 weeks was $ (-1.73, 0.14) $ for ADAS-Cog14 and $ (0.17, 1.83) $ for ADCS-iADL. The sample sizes for both treatment arms and estimations for fixed effects in both treatment arms with standard deviation (SD) or standard error (SE) are documented in Table \ref{table:EXPEDITION3-FIXEDEFFECT}. We use ADCS-iADL to illustrate the assessment of the impact of ETZ variability components on Confirmability.

\begin{table}[H]
\centering
\resizebox{\linewidth}{!}{\begin{tabular}{|l|r|r|r|r|r|r|}
\hline
\multicolumn{1}{|c|}{\multirow{2}[4]{*}{Outcome}} & \multicolumn{2}{c|}{Raw Score at Baseline} & \multicolumn{2}{c|}{Raw Score at 80 Wk} & \multicolumn{2}{c|}{Least-Squares Mean Change at 80 Wk} \\
\cline{2-7}
~& \multicolumn{1}{l|}{Placebo} & \multicolumn{1}{l|}{Solanezumab} & \multicolumn{1}{l|}{Placebo} & \multicolumn{1}{l|}{Solanezumab} & \multicolumn{1}{l|}{Placebo} & \multicolumn{1}{l|}{Solanezumab} \\
\hline
ADCS-iADL score & 45.37$\pm$8.14 & 45.60$\pm$7.93  & 39.01$\pm$11.86 & 39.83$\pm$11.41  & -7.17$\pm$0.32  &-6.17$\pm$0.32\\
\hline
\hline
\multicolumn{1}{|c|}{\multirow{2}[4]{*}{ }} & \multicolumn{2}{c|}{Sample Size at Baseline} & \multicolumn{2}{c|}{Sample Size at 80 Wk} & \multicolumn{2}{c|}{Sample Size for Change at 80 Wk} \\
\cline{2-7}
~& \multicolumn{1}{l|}{Placebo} & \multicolumn{1}{l|}{Solanezumab} & \multicolumn{1}{l|}{Placebo} & \multicolumn{1}{l|}{Solanezumab} & \multicolumn{1}{l|}{Placebo} & \multicolumn{1}{l|}{Solanezumab} \\
\hline
Sample size & 1063 & 1053 & 896 & 908 & 980 & 981\\
\hline
\end{tabular}}
\caption{The sample sizes for both treatment arms and estimations for fixed effects in both treatment arms with standard deviation (SD) or standard error (SE). The plus–minus ($\pm$) values for the scores at baseline and at 80 weeks are means (±SD). The estimated difference is the least-squares mean (±SE) change from baseline between the two trial groups at 80 weeks. The sample size for change is estimated by the average of sample size at baseline and 80 weeks.}
\label{table:EXPEDITION3-FIXEDEFFECT}
\end{table}

\paragraph*{Assessing clinical meaningfulness of treatment effects}
Table 2 in \cite{honig2018trial} gave $ \mathrm{SD}(Y^{Trt[1]}_{i}) $, $ \mathrm{SD}(Y^{Trt[m]}_{i}) $, and $ \mathrm{SE}(Y^{[\mathrm{change}]}_{i}) $ for seven outcome measures, with those for ADCS-iADL reproduced in Table \ref{table:Expedition3VarsIADL} below.

\begin{table}[H]
\centering
\begin{tabular}{|c|c|c|}\hline
 & Visit 1 & Visit $ m = 7$ \\
\hline
Variance & $ \operatorname{Var}\left(Y^{Trt[1]}_{i}\right)=64.580$ & $\operatorname{Var}\left(Y^{Trt[m]}_{i}\right)=135.389$ \\
\hline
Standard Deviation & $8.036$ & $11.636$ \\
\hline
\end{tabular}

\vspace{0.5em}

\begin{tabular}{|c|c|}\hline
 & Change \\
\hline
Variance & $\operatorname{Var}\left(Y^{[\mathrm{change}]}_{i}\right)=92.365$ \\
\hline
Standard Deviation & $9.611$ \\
\hline
\end{tabular}
\caption{Reported EXPEDITION3 ADCS-iADL variances. Visit 1 and visit 7 variances are computed from raw scores. Variance for Change is converted from the reported SE for LSmeans of Change using Hedges' formula.}
\label{table:Expedition3VarsIADL}
\end{table}

Applying ETZ decomposition, we obtain the estimated ETZ ADCS-iADL variability components in Table \ref{table:Expedition3ETZvariances}.

\begin{table}[H]
\centering
\begin{tabular}{|c|c|c|}\hline
Random Intercept & Measurement Error & Random Trajectory \\
\hline
$53.802$ & $10.778$ & $70.809$ \\
\hline
\end{tabular}

\vspace{0.5em}

\begin{tabular}{|c|c|c|}\hline
SD(Z) & SD(E) & SD(Traj) \\
\hline
$7.335$ & $3.283$ & $8.415$ \\
\hline
\end{tabular}
\caption{Estimated ETZ ADCS-iADL EXPEDITION3 variability components.}
\label{table:Expedition3ETZvariances}
\end{table}

Having an estimate of the within patient SD of measurement error is useful for patients and clinicians to interpret the estimated effect of a specific treatment. That is, when interpreting the average decrease of ADCS-iADL over an 80-week period reported in \cite{honig2018trial} being $ -7.17 $ for patients given the placebo and $ -6.17 $ for patients given solanezumab, one should keep the 3.283 SD of measurement error in mind.

Having an estimate of the between patient SD of trajectory is useful for patients and clinicians to interpret the estimated \textit{difference} of treatment effects. That is, when interpreting the average \textit{difference} of Change in ADCS-iADL between solanezumab and placebo reported in \cite{honig2018trial} being $ 1.0 $, one should keep the 8.415 SD of trajectory in mind.

\subsection{Confident Efficacy calculation}
Denote true efficacy $ \mu^{[m]}_{Rx}+\mu^{[m]}_{C}-\mu^{[1]}_{Rx}-\mu^{[1]}_{C}$ by $\theta_{Phase 2}$, and denote its point estimate from Phase 2 data by $\Bar{\theta}_{Phase 2}$. With $SE_{pooled}$ being the pooled standard error of \textit{change},
\[
\frac{\Bar{\theta}_{Phase 2} - \theta_{Phase 2}}{SE_{pooled}} \sim t_{n_{Rx}+n_C-2},
\]
where $n_{Rx}$ and $n_C$ are the $Rx$ and $ C $ sample sizes.

There are two types of original outcome measures. ADAS-Cog and HbA1c are examples of lower outcome being better, while ADL is an example of higher outcome is better. We take the type of higher outcome being better as example to illstrate and the other type is similar. Our Confident Efficacy, denoted by $L_{Phase2}$, is the $ t $ distribution lower $1-(d_{Phase2}+50\%)$ confidence limit for $\theta_{Phase 2}$ 
\[
\Bar{\theta}_{Phase 2} - t_{d_{Phase2}+50\%,n_{Rx}+n_C-2}\times SE_{pooled}.
\]
Different Phase 2 studies independently test different compounds for different diseases. By Kolmogorov's Law of Large Numbers (LLN) for independent but not necessarily identically distributed random variables (Theorem D on page 27 of \cite{Serfling(1980)}), there is confidence that the proportion of Phase 2 studies with $L_{Phase2}$ higher than its true $\theta_{Phase 2}$ is $1-(d_{Phase2}+50\%)$.

\subsection{Confidently Bounded Quantile calculation}
The point estimate of the efficacy $\Bar{\theta}_{Phase 3 \mid Phase 2}$ from a Phase 3 study for $\theta_{Phase 3 \mid Phase 2}$, should the sponsor decide to transition to it, is assessed as follows. $\Bar{\theta}_{Phase 3 \mid Phase 2}$ is assumed to follow the Normal distribution with mean $L_{Phase2}$ and variance $Var (\Bar{\theta}_{Phase 3 \mid Phase 2})$. $L_{Phase2}$ is taken as the mean of $\theta_{Phase 2}$, with
\[
Var (\Bar{\theta}_{Phase 3 \mid Phase 2}) = \sigma^2_{pooled}\sqrt{\frac{1}{n_{Phase 3,Rx}}+\frac{1}{n_{Phase 3,C}}},
\]
where $\sigma^2_{pooled} = Var(Y^{[\mathrm{change}]r}_{Rx}) = Var(Y^{[\mathrm{change}]r}_{C}) = Var(Traj^{Trt[m]}_{i}) + 2 \times Var(E)$, $n_{Phase 3,Rx}$ and $n_{Phase 3,C}$ are sample sizes set for $Rx$ and $C$ in Phase 3, respectively. Then, as illustrated in Figure \ref{fig:Combinedplot}, a histogram of potential $ \Bar{\theta}_{Phase 3 \mid Phase 2} $ is obtained by replicating the Phase 3 study many times with the sponsor's chosen sample size $n_{Phase 3,Rx}$ and $n_{Phase 3,C}$, generating $ \Bar{\theta}_{Phase 3 \mid Phase 2} $ from a Normal distribution. The Confidently Bounded Quantile $L_{Phase3\mid Phase 2}$ is then the $1-(d_{Phase3 \mid Phase2}+50\%)$ quantile of this distribution. Conditional on $L_{Phase2}$ being the true efficacy $\theta_{Phase 2}$, the proportion of replicated Phase 3 studies with outcome higher than $L_{Phase 3 \mid Phase 2}$ is $1-(d_{Phase3 \mid Phase2}+50\%)$. The uncertainty in estimating $\theta_{Phase 3 \mid Phase 2}$ reduces when sample sizes increase and the Confidently Bounded Quantile changes as sample sizes change.

\subsection{Making decision with variabilities as they are, an example}
The \textit{Transition Decision-Making Process Plots} tab of our App calculates the confident quantile with estimated separation and variability from the feeder study ``as is''. We use EXPEDITION3 as a feeder study to illustrate how this facilitates the decision-making process. The outcome considered is ADCS-iADL, which is better when it is higher. Before decision is made, two requirements are set below:
\begin{itemize}[leftmargin=1.4em]
\item requires the probability that Phase 2 study transitions to Phase 3 study correctly is at least, for example, 95\%, and
\item requires the probability that Phase 3 study shows efficacy conditioned on that Phase 2 has transitioned to Phase 3 study correctly is at least, for example, 80\%.
\end{itemize}
In this case, the success confidence $95\% \times 80\% = 76\%$ is guaranteed. The two requirements are achieved by \textbf{Confident Efficacy} and \textbf{Confidently Bounded Quantile}, and visualized in Figure \ref{fig:Combinedplot}.

Referring to the concept of Confident Efficacy proposed in Section \ref{sec:DMProcess} and the estimations from Table \ref{table:EXPEDITION3-FIXEDEFFECT}, we can calculate the Confident Efficacy as
\[
L_{Phase2} = \Bar{\theta}_{Phase 2} - t_{1-(d_{Phase2}+50\%),n_{Rx}+n_C-2}\times SE_{combined}
= 1 - t_{95\%,1959}\times0.32 = 0.26.
\]
Similarly, we construct Confidently Bounded Quantile which is \textit{conditional} quantile $L_{Phase 3\mid Phase 2}$ for Phase 3 study. It is conditional because it constructs on the distribution of efficacy in Phase 3 conditioned on Phase 2 study including discount.

Referring to Section \ref{sec:DMProcess} and setting the sample size for each arm ($Rx$ and $C$) in Phase 3 as 1000, the Confidently Bounded Quantile (CBQ) can be constructed as
\begin{align*}
L_{\text{Phase 3}\mid \text{Phase 2}} 
&= L_{\text{Phase 2}} 
- Z_{d_{\text{Phase 3} \mid \text{Phase 2}} + 50\%} 
\times \sigma_{\text{pooled}}
\sqrt{\frac{1}{n_{\text{Phase 3,Rx}}} + \frac{1}{n_{\text{Phase 3,C}}}} \notag\\
&= 0.26 
- Z_{80\%} \times 
\sqrt{70.809 + 2\times 10.778} 
\times \sqrt{\frac{1}{1000}+\frac{1}{1000}} 
= -0.1.
\end{align*}
EXPEDITION3 indeed failed in real life in 2016. Following CBQ proposed, $L_{Phase 3\mid Phase 2}$ calculated is negative, which means that the predefined success confidence cannot be guaranteed and consists with the result of EXPEDITION3. However, if sample size for each arm increases to 2000, the CBQ becomes 0.1 (positive), which may lead to the success of EXPEDITION3.

\section{How decisions change if variabilities are changed}\label{sec:ComponentImpacts}
Given data from a feeder study, the \textit{Profile Plots} tab of our Confirmability app assesses how much each of the variability reduction strategies described in Section \ref{sec:ReducingVar} can impact on the probability of having a successful Confirmatory study. We use real data examples to show
\begin{itemize}[leftmargin=1.4em]
\item $ Var (Z) $ hardly impact confirmability if effects are measured as \textit{change-from-baseline}.
\item Change in $ Var (E) $ affects the \textit{shape} of trajectory.
\item Change in $ Var (Traj) $ affects \textit{separation} at the milestone visit.
\end{itemize}
After Section \ref{sec:TrulicityEntry} shows variability of $ Z $ hardly impacts Var(Change), Section \ref{sec:EXPEDITION3confirmability} shows how our Confirmability App reacts instantaneously to ``what happens to a Phase 3 study if $ Var (Traj) $ or $ Var (E) $ changes'' questions by visualizing the impact using profile plots. A sponsor can then decide whether to invest in one or more Section \ref{sec:ReducingVar} strategies to reduce the variability of impactful component(s), to increase the chance of \textit{confirmability}.

\subsection{Baseline variability impacts Var(Change) little}\label{sec:TrulicityEntry}
There is the thinking that narrowing the patient entry criterion makes patients more homogeneous and thereby makes clinical trials more likely to succeed. This turns out not to be the case if treatment effects are measured in terms of Change, because (\ref{eq:VarChange}) in the ETZ decomposition shows SD(Change) does not involve patients' baseline measurements $ Z $. We demonstrate this using information available in the FDA's review of Trulicity\textsuperscript{\textregistered} for treating patients with type 2 diabetes.

Trulicity\textsuperscript{\textregistered} (compound name dulaglutide) is a once-weekly injectable prescription medicine to improve blood sugar (glucose) in adults with type 2 diabetes (mellitus), as measured by control of haemoglobin A1c (HbA1c). Biologic License Application 125469 for Trulicity/dulaglutide had five studies. Patient entry criteria, SD(Basline), and SD(Change) for studies GBDA and GBDC, reported on pages 6 and 7 in \cite{TrulicityFDAsumaryReview(2014)} and on pages 58 and 61 of \cite{TrulicityFDAstatisticalReview(2014)}, are reproduced in Tables \ref{table:TrulicityRewindGBDA} and \ref{table:TrulicityRewindGBDC}.

\begin{table}[H]
\centering
\begin{tabular}{|c|c|c|c|}
\hline
HbA1c & Dulaglutide + Metformin + Pioglitazone & Placebo & Exenatide \\
\hline
$ n $ & 279 & 141 & 276 \\
\hline
SD(Baseline) & 1.3 & 1.3 & 1.3 \\
\hline
SD(Change) & 0.98 & 0.97 & 1.02 \\
\hline
\end{tabular}
\caption{GBDA sub-study: Dulaglutide + Metformin + Pioglitazone vs. Placebo and Exenatide with baseline HbA1c $ \in [7, 11] $}
\label{table:TrulicityRewindGBDA}
\end{table}

\begin{table}[H]
\centering
\begin{tabular}{|c|c|c|}
\hline
HbA1c & Dulaglutide & Metformin \\
\hline
$ n $ & 269 & 268 \\
\hline
SD(Baseline) & 0.9 & 0.8 \\
\hline
SD(Change) & 1.00 & 0.98 \\
\hline
\end{tabular}
\caption{GBDC sub-study: Dulaglutide vs. Metformin with baseline HbA1c $ \in [6.5, 9.5] $}
\label{table:TrulicityRewindGBDC}
\end{table}

As can be seen, narrowing the HbA1c baseline entry criterion from [7, 11] in GBDA to [6.5, 9.5] in GBDC decreases SD(baseline) of HbA1c from 1.3 to within [0.8, 0.9], but hardly affects SD(Change), which remains steady within [0.97, 1.02] for both studies. So decreasing variability of intercept $ Z $ may not enhance confirmability much. Thus we turn our attention to how variability in trajectory and measurement error impact confirmability.

Depending on outcome measures, response profiles can appear to have different shapes. For example, from Figure \ref{fig:SolanezumabProfiles}, in terms of Change, while ADCS-iADL response profiles appear close to be linear, response profiles of ADAS-Cog seem not to be easily parameterized. However, as we will show in Figure \ref{fig:CombinedProfilePlot} of Section \ref{sec:EXPEDITION3confirmability}, variability of measurement error $ E $ can distort the shape of response profiles, by making linear response profiles appear non-linear for example. To understand how variability of the trajectory and measurement error separately impact observed separation between $ Rx $ and $ C $ at the end of the study and shape of the response profiles, we connect our ETZ modeling principle with the so-called random coefficients model.

\subsection{Assessing the replicability of a Confirmatory study}\label{sec:RandomCoefficient}
Calculation under the \textit{Transition Decision-Making Process Plots} tab does not assume any functional shape for response profiles so that, if the calculated quantile exceeds the desired threshold, one can be confident it is so without such assumptions. However, if the calculated quantile fails to exceed the desired threshold, then one can turn to the \textit{Profile Plots} tab of our App to see whether investing in some strategies in Section \ref{sec:ReducingVar} might make the Confirmatory study more likely to succeed.

The \textit{Profile Plots} tab allows the user to change each of the ETZ variability components, and assess \textit{replicability} of a Confirmatory study by observing how stable or unstable the response profiles are upon replications of the study, proceeding only if they are stable. Simulated replications of a Confirmatory study under the \textit{Profile Plots} tab assumes the response profiles have idealized functional forms (a confirmatory study not replicable under an ideal situation will not be replicable under non-ideal situations). For our proof-of-concept App, at present calculations under the \textit{Profile Plots} tab are for straight line response profiles, leaving development of more sophisticated profiles for future research.

Corresponding to the RANDOM statement in Proc Mixed of SAS, a random coefficients model considers Time as a continuous variable, with the response of each subject being generated randomly from a distribution of response trajectories as a function of Time (e.g., random intercept and random slope with a bivariate Normal distribution). Expectations of the response trajectories are the response profiles, reflecting the average effects of treatments $ Rx $ and $ C $ over Time.

In contrast to MMRM, but similar to the ETZ modeling principle, a random coefficients model has, in addition to randomness in the intercepts and the response trajectories, a residual term in the model which is analogous to our measurement error term. With treatments indexed by $ Trt = Rx \mbox{ or } C $, a random coefficients model for linear response profiles would be
\begin{eqnarray}\label{model:RandomCoefficients}
Y_{i}^{Trt[v]} = \alpha^{Trt} + a_{i}^{Trt} + \beta^{Trt} x_{i}^{[v]} + b_{i}^{Trt} x_{i}^{[v]} + e_{i}^{Trt[v]},
\end{eqnarray}
where $ x_{i}^{[v]} $ is the Time at which the $ v^{th} $ measurement on the $ i^{th} $ subject is taken. The fixed effects are represented by $ \alpha^{Trt} + \beta^{Trt} x_{i}^{[m]} $. What distinguishes the ETZ modeling principle from this (general) random coefficient model (\ref{model:RandomCoefficients}) is $ \alpha^{Rx} = \alpha^C $ in (\ref{eq:EqualAlpha}) because at Time = 0 ($ x_{i}^{[1]} \equiv 0 $) patients have yet to be treated.

Random subject effect for the $ i^{th} $ patient assigned to treatment $ Trt $ is $ (a_{i}^{Trt}, b_{i}^{Trt})' $, with
\[
\left[ \begin{array}{c}
a_{i}^{Trt} \\
b_{i}^{Trt}
\end{array} \right] \sim i.i.d. ~ MVN \left( 
\left[ \begin{array}{c}
0 \\
0
\end{array} \right] 
, 
\left(
\begin{array}{cc}
\sigma_a^2 & \sigma_{ab} \\
\sigma_{ab} & \sigma_b^2
\end{array}
\right)
\right).
\]
The residuals $ e_{i}^{Trt[v]} $ are assumed to be i.i.d. $ N(0, \sigma^2) $, and independent of $ (a_{i}^{Trt}, b_{i}^{Trt})' $. Thus, with $ t^{[m]} $ denoting time of the milestone visit (visit $ m $),
\begin{eqnarray}
Var(Y_{i}^{Trt[1]} | x_{i}^{[1]} = 0) & = & \sigma_a^2 + \sigma^2\\
Var(Y_{i}^{Trt[m]} | x_{i}^{[m]} = t^{[m]}) & = & \sigma_a^2 + 2 t^{[m]} \sigma_{ab} + {t^{[m]}}^2 \sigma_b^2 + \sigma^2 \\
Var(Y_{i}^{Trt[\mathrm{change}]}) & = & {t^{[m]}}^2 \sigma_b^2 + 2 \sigma^2.
\end{eqnarray}
Randomness of the intercept, response trajectories, and residuals, together determine the $ m \times m $ variance-covariance matrix of outcome measures on the subjects across the $ m $ visits.

For a linear random coefficients model, our assumption of independence between $ Z $ and $ Traj $ would be $\sigma_{ab} = 0$. In practice, a user can assess closeness of $\sigma_{ab} = 0$ to zero in a variety of ways to feel comfortable proceeding with this assumption. Matching with (\ref{eq:VarZest}), (\ref{eq:VarEest}), and (\ref{eq:VarChange}) then, we see that in ETZ notations
\begin{eqnarray}
\begin{bmatrix}
a_{i}^{Trt} \\ b_{i}^{Trt}
\end{bmatrix}
& \sim i.i.d. ~ & MVN
\left(
\begin{bmatrix}
0 \\ 0
\end{bmatrix}, 
\begin{bmatrix}
Var(Z) & 0 \\ 0 & Var(Traj)/{t^{[m]}}^2
\end{bmatrix}
\right)  \label{eq:BivariateDist} \\
e_{i}^{Trt[v]} & \sim i.i.d. ~ & N(0, Var(E^{[m]}_{i})). \label{eq:ResidualDist}
\end{eqnarray}
Note that, while intercepts are in units of $ Y $ measurements (e.g., ADAS-Cog), with Time being a continuous variable in the random coefficient setting, slopes $ \beta^{Trt} $ and $ b_i $ are in units of $ Y $ measurements divided by units of time (e.g., ADAS-Cog per week, if Time is measured in weeks).

For illustration purpose, we pretend EXPEDITION3 is the feeder study to itself, to see to what extent a similar sample size (2000+ patients) feeder study might replicate/confirm its own finding in ADCS-iADL. Assuming true response profiles are linear, random coefficient model (\ref{model:RandomCoefficients}) is used for simulation. (Fixed) treatment effects are estimated from visit 1 (week 0) and visit 7 (week 80) point estimates reported in \cite{honig2018trial}, with
\[
\alpha^{Rx} = 45.6,\quad \beta^{Rx} = -\frac{6.17}{80},\qquad
\alpha^{C}  = 45.37,\quad \beta^{C} = -\frac{7.17}{80}.
\]
Then, assuming the two treatment arms share a same covariance matrix, each simulated study uses a different random number seed to generate data with initial sample sizes same as in EXPEDITION3 and no patient dropout, with
\[
\begin{bmatrix}
a^{Trt}\\b^{Trt}
\end{bmatrix}  \sim MVN\left(
\begin{bmatrix}
0\\0
\end{bmatrix}, 
\begin{bmatrix}
53.802&0\\0&70.809/t_m^2
\end{bmatrix}\right),
\]
and i.i.d. measurement error $e^{Trt[v]} \sim N(0,10.778)$, where $Trt = Rx$ or $C$.

\subsection{Impact of \texorpdfstring{$\operatorname{Var}(Traj)$ and $\operatorname{Var}(E)$}{Var(Traj) and Var(E)} on a Phase~3 study}\label{sec:EXPEDITION3confirmability}
\begin{figure}[htbp]
\centering
\includegraphics[width=0.8\linewidth]{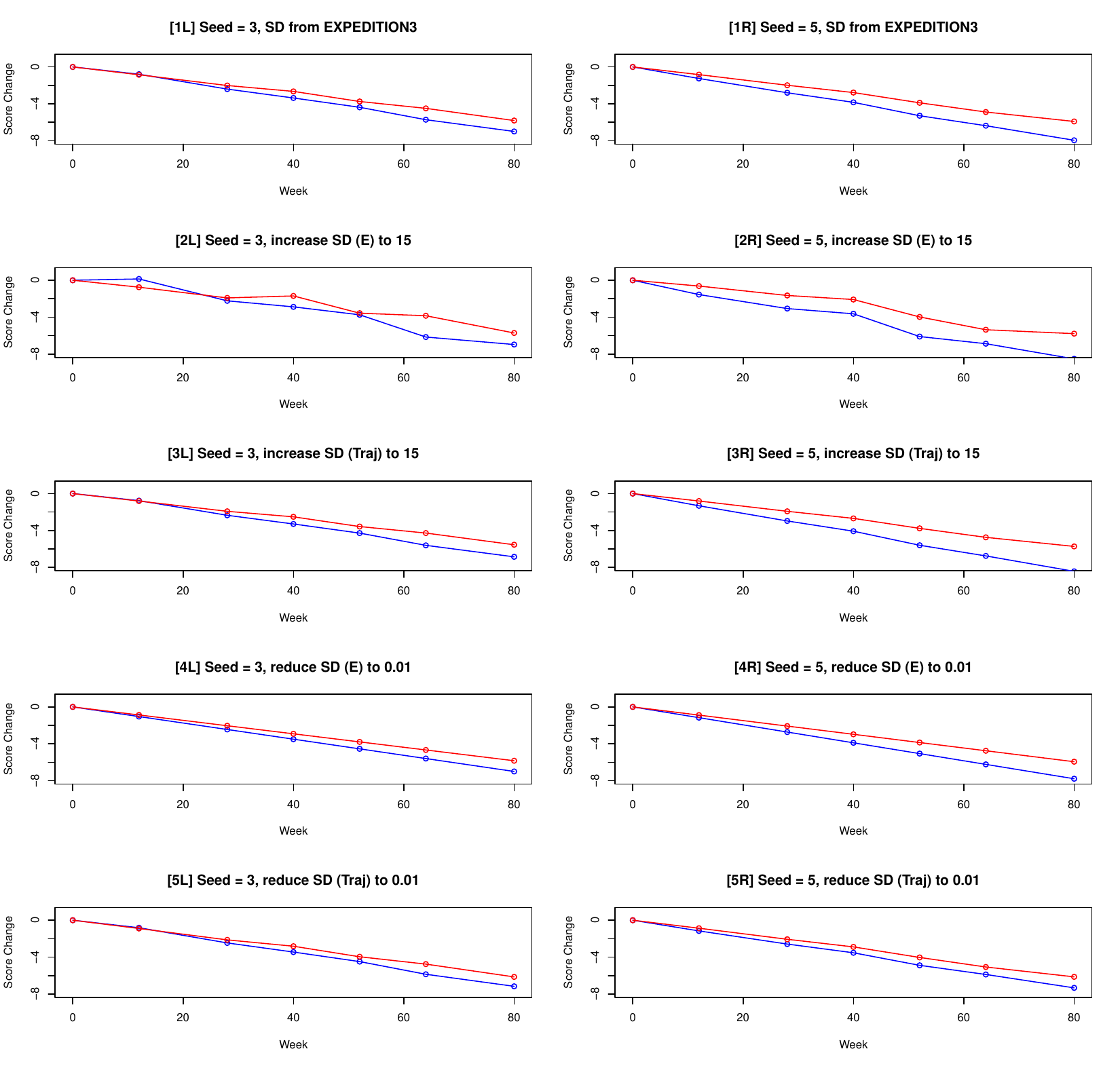}
\caption{Changing the random number Seed (from 3 in the left column to 5 in the right column as an example) triggers the \textit{Profile Plots} tab of our Confirmability App to simulate an independent replication of the Confirmatory study. With means and ETZ variability components as estimated for EXPEDITION 3 in Tables \ref{table:EXPEDITION3-FIXEDEFFECT} and \ref{table:Expedition3ETZvariances}, [1L] \& [1R] show separation at Week 80 lacks stability. Increasing $ Var(E) $ and $ Var(Traj) $ each separately while keeping the other as in Table \ref{table:Expedition3ETZvariances}, [2L] \& [2R] show large $ SD(E) $ distorts the shape of the profiles, while [3L] \& [3R] show large $ SD(Traj) $ causes separation at the milestone to be unstable. With $ SD(Traj) = 8.415 $ still as estimated for EXPEDITION 3, separation remains unstable even with $ Var(E) $ reduced to 0.1, as [4L] \& [4R] show. With $ SD(E) = 3.283 $ as estimated for EXPEDITION 3, reducing $ SD(Traj) $ to 0.1 does make the $ Rx $ and $ C $ separation at week 80 less variable, as [5L] \& [5R] show.}
\label{fig:CombinedProfilePlot}
\end{figure}

The \textit{Profile Plots} tab of our App shows interactively what happens to a Phase 3 study if any of the three ETZ variability components changes. Taking EXPEDITION3 as an example feeder study, Figure \ref{fig:CombinedProfilePlot} gives examples of changing profile plots in a single Phase 3 study if $ Var(E) $ or $ Var(Traj) $ changes. We observe from Figures (4L) and (4H) that reducing $Var(E)$ alone is insufficient to ensure stability of separation at the milestone. For EXPEDITION3, we observed from Figures (5L) and (5H) that reducing $ Var(Traj) $ likely will have a bigger impact in enhancing Replicability of the Confirmatory study.

\paragraph*{Recommended strategies}
For targeted therapies, a potential strategy to reduce $Var(Traj)$ is \textit{Subgroup Targeting}, enrolling only patients with sufficient drug target expression. For Alzheimer's disease medicine whose mechanism of action is to reduce amyloid beta plaques, a sponsor may consider enrolling only patients with substantial amyloid beta deposition.

Using a \textit{Surrogate Endpoint} to measure outcome is a strategy that potentially reduces both $Var(Traj)$ and $Var(E)$, but its benefit requires solid biomedical and statistical justifications. For example, in using HbA1c measurement as a surrogate to microvascular events in T2DM, there is epidemiological evidence that controlling HbA1c reduces microvascular complications associated with diabetes mellitus, and its $Var(E)$ is small (around 0.03).

\section{Scope, limitation, key messages, and future research}
The ETZ modeling principle is not only applicable to the major disease areas in Table \ref{table:NofPatients}, but also to rare disease areas. Of the some 7000 rare diseases, fewer than 10\% have approved treatments. The challenge is large variability of functional outcome measures coupled with a small number of patients with rare diseases makes the identification of reliable clinical outcome assessments (COAs) for efficacy endpoints difficult. ETZ's ability to suggest judicious allocation of resources to reduce sources of variability should be useful for COA research.

\begin{description}[leftmargin=1.2em]
\item[Scope of application] Applicable to studies with Before-and-after treatment Repeated Measurements, the ETZ modeling principle does not assume a particular shape for the response profiles.
\item[Limitation] However, the assessment of replicability of estimated response profiles does assume a shape for the response profiles (linear profiles in the current implementation).
\item[Key messages]
\begin{itemize}
\item Efficacy measured in terms of \textit{change} (from baseline) is largely unaffected by range of the patient entry criterion.
\item Large \textit{measurement error} variability will (even with large sample sizes) cause separation between $ Rx $ and $ C $ at the milestone visit to be unstable; and linear response profiles to appear non-linear.
\item Large \textit{trajectory} variability will cause separation between $ Rx $ and $ C $ at the milestone visit to not be replicable (even with large sample sizes).
\end{itemize}
\end{description}

\section{Interest statement}
Y.S., Y.H., X.W., S-Y.T. declare no competing interests. Y. L. is an employee and shareholder of Eli Lilly. J.C.H. is a professor emeritus of the Ohio State University and a consultant to Eli Lilly.

\bibliographystyle{Chicago}
\bibliography{Xreference}
\end{document}